\documentclass[sort&compress]{elsarticle}

\journal{Physica D}

\usepackage{amsmath} 
\usepackage{amssymb} 
\usepackage{bm} 

\usepackage{tikz}   
\usetikzlibrary{arrows}  

\newcommand{\bx}{{\bm x}}
\newcommand{\bk}{{\bm k}}
\newcommand{\bq}{{\bm q}}
\newcommand{\bu}{{\bm u}}
\newcommand{\bJ}{{\bm J}}
\newcommand{\cL}{{\mathcal L}}
\newcommand{\cLdag}{{\mathcal L}^{\dagger}}
\newcommand{\cO}{{\mathcal O}}
\newcommand{\bv}{{\bm v}}
\newcommand{\bvdag}{{\bm v}^{\dagger}}
\newcommand{\bff}{{\bm f}}
\newcommand{\bg}{{\bm g}}

\newcommand{\DUi}{{D_{U_1}}}
\newcommand{\DVi}{{D_{V_1}}}
\newcommand{\DUj}{{D_{U_2}}}
\newcommand{\DVj}{{D_{V_2}}}

\newcommand{\Qzh}{Q_{zh}}
\newcommand{\Qzz}{Q_{zz}}
\newcommand{\Qzw}{Q_{zw}}
\newcommand{\Qwh}{Q_{wh}}
\newcommand{\Qww}{Q_{ww}}
\newcommand{\Qwz}{Q_{wz}}

\newcommand{\bNLT}{{\textbf{NLT}}}

\newcommand{\phm}{\phantom{-}} 
\newcommand{\phz}{\phantom{0}} 

\newcommand{\etal}{\hbox{\emph{et al.}}}

\newcommand{\red}{\color{black}}

\def\clap#1{\hbox to 0pt{\hss#1\hss}}

\def\mathrlap{\mathpalette\mathrlapinternal}

\def\mathrlapinternal#1#2{%
\rlap{$\mathsurround=0pt#1{#2}$}}

\begin{document}

\begin{frontmatter}

\title{Spatiotemporal chaos and quasipatterns in coupled reaction--diffusion systems}

\author[LeedsMathsaddress,LeedsCompaddress]{Jennifer K. Castelino}
\author[Loughboroughaddress]{Daniel J. Ratliff}
\author[LeedsMathsaddress]{Alastair M. Rucklidge\corref{cor1}}
\author[LeedsMathsaddress,Oxfordaddress]{Priya Subramanian}
\author[Williamsaddress]{Chad M. Topaz}

\address[LeedsMathsaddress]{School of Mathematics, University of Leeds, Leeds, LS2 9JT, UK}
\address[LeedsCompaddress]{School of Computing, University of Leeds, Leeds, LS2 9JT, UK}
\address[Loughboroughaddress]{Department of Mathematical Sciences, Loughborough University, Loughborough, LE11 3TU, UK}
\address[Oxfordaddress]{Mathematical Institute, University of Oxford, Oxford, OX2 6GG, UK}
\address[Williamsaddress]{Department of Mathematics \& Statistics, Williams College, Williamstown MA, 01267, USA}

\cortext[cor1]{Corresponding author}

 \begin{abstract}
In coupled reaction--diffusion systems, modes with two different length scales can interact to produce
a wide variety of spatiotemporal patterns. Three-wave interactions between these modes
can explain the occurrence of spatially complex steady patterns and time-varying states
including spatiotemporal chaos. The interactions can take the form of two short waves
with different orientations interacting with one long wave, or vice verse. We investigate
the role of such three-wave interactions in a coupled Brusselator system. As well as
finding simple steady patterns when the waves reinforce each other, we can also find
spatially complex but steady patterns, including quasipatterns. When the waves compete with each
other, time varying states such as spatiotemporal chaos are also possible. The signs of the
quadratic coefficients in three-wave interaction equations distinguish between these two cases. By
manipulating parameters of the chemical model, the formation of these various states can be
encouraged, as we confirm through extensive numerical simulation. Our arguments allow us to predict
when spatiotemporal chaos might be found: standard nonlinear methods fail in this case. The
arguments are quite general and apply to a wide class of pattern-forming systems, including the
Faraday wave experiment.
 \end{abstract}

\begin{keyword}
Turing patterns, Brusselator, three-wave interactions, quasipatterns,
spatiotemporal chaos
\MSC[2010] 35B36\sep 70K55\sep 35K57\sep 37L99\sep 70K30.
\end{keyword}

\end{frontmatter}


\section{Introduction}

\label{sec:intro}

Two substances that react and diffuse can form patterns, an insight first highlighted in the work of Alan Turing~\cite{Tur1952,Dawes2016}. Motivated by an interest in embryonic morphogenesis, Turing studied discrete and continuum models for the spontaneous emergence of structure in a ring of cells. Depending on the details of the reaction, a Turing-type system may have a stable, spatially-uniform steady state in the absence of diffusion. Two fundamental instabilities may occur. One possibility is a Hopf bifurcation leading to temporal oscillations with a preferred wavenumber of zero. The other possibility, driven by diffusion, is a bifurcation to a steady spatial pattern (Turing pattern) with non-zero wavenumber, typically stripes or hexagons.

The first laboratory experiment to produce a Turing pattern came nearly 40 years after Turing's original work: Ref.~\cite{CasDulBoi1990} reports the observation of patterns in the chlorite--iodide--malonic acid (CIMA) chemical reaction. Since the seminal discoveries of~\cite{Tur1952,CasDulBoi1990}, there has been a vast literature on reaction--diffusion patterns and their applications, which include animal skin pigmentation~\cite{NakTakKan2009}, the cerebral cortex~\cite{Car2002}, vegetation ecology~\cite{Kla1999}, plankton colonies~\cite{LevSeg1976}, and many others~\cite{KonMiu2010,MaiWooBak2012}.

A variation on the classic reaction--diffusion system is the so-called coupled (or multilayered) system, in which two or more reaction--diffusion systems are connected together so that they may influence each other. Because of the additional degrees of freedom in these coupled systems, they are an amenable setting in which to investigate how competing instabilities affect pattern formation. Coupled systems can produce a variety of states including simple Turing patterns, standing waves, mixes of Turing patterns and spiral waves, square and hexagonal superlattice patterns, and many more~\cite{YanDolZha2002,YanDolZha2006}. Coupled systems are important in biology, especially in neural, ecological and developmental contexts; see~\cite{EpsBerDol2008} for examples and for an overview of selected results. 

\begin{figure}
 \hbox to \hsize{%
  \includegraphics[width=0.55\hsize]{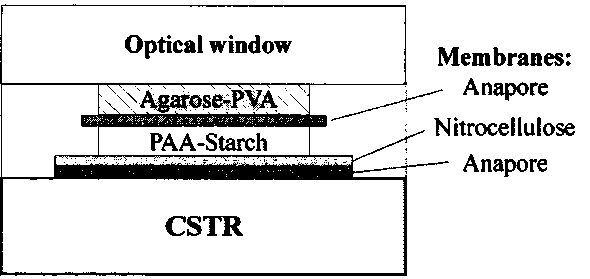}\hfill
  \includegraphics[width=0.44\hsize]{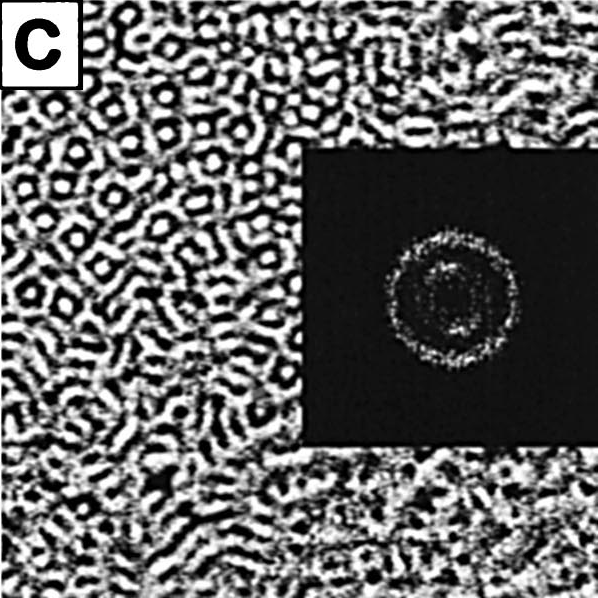}}
\caption{Patterns formed when the chlorine dioxide--iodine--malonic acid (CDIMA) reaction
occurs in two layers that are diffusively coupled. Left: Experimental setup: the chemical 
reactions occur in the agarose--PVA and PAA--starch layers, separated by an Anapore 
membrane. Right:
Patterns with two different length-scales ($0.46\,$mm and $0.25\,$mm) visualised 
through a filter transparent to red light 
(which highlights the pattern in the PAA--starch layer),
with the corresponding spatial power spectrum
in the inset.
Reproduced from~\cite{BerDolYan2004}
with permission.}
\label{fig:Berenstein}
\end{figure}

Unfortunately, it is difficult to manipulate experiments on the aforementioned biological systems. A common approach to studying coupled reaction--diffusion systems, then, is to study a paradigmatic chemical experiment in the laboratory, as in~\cite{BerDolYan2004} (see Figure~\ref{fig:Berenstein}). In this work, the experimentalists set up two thin gels, within each of which the chlorine dioxide--iodine--malonic acid (CDIMA) reaction takes place. They put the gels in contact and controlled the strength of coupling between the two layers by modifying the properties of a membrane placed at the interface, resulting in different patterns. To complement laboratory experiments, investigators have studied a host of nonlinear partial differential equation (PDE) models, including the Lengyel--Epstein model of the CIMA reaction~\cite{LenEps1991} and Brusselator model of a generic trimolecular reaction~\cite{PriLef1968}.

The work of~\cite{CatMcNTop2012} included a theoretical study of two reaction--diffusion systems coupled together in a parameter regime near a codimension-two Turing--Turing bifurcation point. This work demonstrated that by changing the interlayer coupling strength, one can manipulate the ratio of the length scales associated with two resonantly interacting Turing instabilities and encourage the formation of certain complex patterns in the Brusselator model. In our present work, we will also carry out a theoretical investigation of coupled reaction--diffusion systems and we will also focus on resonant mode interactions. However, in contrast to the set-up in~\cite{CatMcNTop2012}, we will use the within-layer diffusion constants as control parameters. In experiments, one could manipulate these diffusion constants by changing properties of the medium of each layer.

Our work complements a robust literature that has examined the role of three-wave interactions in the Faraday system, in which a layer of fluid is vertically vibrated in a time-periodic fashion, potentially producing standing wave patterns. Patterns with two dominant length scales, including quasipatterns and superlattice patterns, have been observed in many Faraday wave experiments~\cite{EdwFau1994,KudPieGol1998,EpsFin2004,DinUmb2006,EpsFin2006}. The theory of Faraday three-wave interactions was developed in~\cite{SilTopSke2000,TopSil2002,PorTopSil2004,TopPorSil2004,SkeRuc2015}, among other sources. Much of this body of work took the following approach. Based on symmetry considerations, one can write down amplitude equations describing the slow-time evolution of modes close to a codimension-two point where all waves associated with two different length scales are neutrally linearly stable. By detuning from that point and assuming that one of the sets of waves is weakly damped, one can perform a centre manifold reduction and assess the role that the weakly damped mode has on the dynamics of the other modes. At a granular level, this influence is seen as a (potential) contribution to coefficients of cubic terms in the amplitude equations for the primary pattern modes. The leading order influence is determined by quadratic terms in the original amplitude equations.

Our present study focuses on the role of three-mode or three-wave interactions
and, pivotally, builds on, clarifies and extends the main ideas
of~\cite{RucSilSke2012}. When there are two (nearly) critical length scales
that are not too disparate, two of the shorter wavelength modes with different
orientations can interact with one of the longer ones, or two of the longer
wavelength modes can interact with one of the shorter ones. {\red In each case,
the orientations of the modes are determined by the requirement that two longer
wavevectors add up to a shorter one, or that two shorter      
wavevectors add up to a longer one.} Pattern formation can be
strongly dominated by these interactions. Rather than slaving away one set of
critical modes and studying cubic terms, as described above, we instead see how
much understanding may be gleaned by restricting our attention to quadratic
terms near the codimension-two point. This approach, namely, studying the
effect of three-wave interactions on spatiotemporal pattern formation in
reaction--diffusion systems by looking at quadratic coefficients, has proven
successful in the past~\cite{RucSilSke2012,VerDeWDew1992}. Our present work
develops a more exhaustive investigation in the context of layered Turing
systems, though the ideas are applicable wherever a pattern-forming system can
have two unstable length scales, including the Faraday wave experiment.

The rest of this paper is organised as follows. In Section~\ref{sec:3WI} we outline the
basic nonlinear three-wave interactions in the case of pattern formation with two
competing wavelengths, and in Section~\ref{sec:quadraticcoefficients} discuss the role of
the quadratic coefficients (and in particular their signs) in influencing the resulting
patterns. Section~\ref{sec:Brusselator} presents the two-layer Brusselator model, and
Sections~\ref{sec:linear} and~\ref{sec:wnlt} describe the linear and weakly nonlinear
theory of the model. Numerical results appear in Section~\ref{sec:numerical}, and we
conclude in Section~\ref{sec:discussion}.


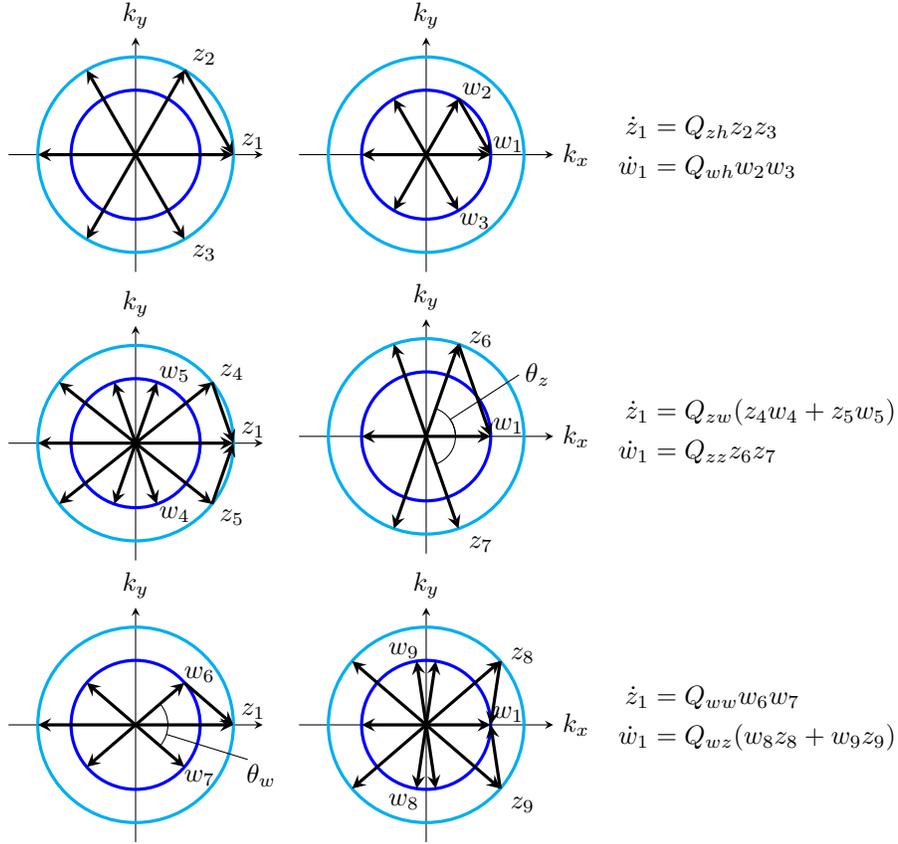
\begin{figure}

%
\hbox to \hsize{%
\begin{minipage}{0.60\linewidth}
%
%
\begin{tikzpicture}[scale=1.30,>=stealth]
   \draw[->] (-1.3,0) -- (1.3,0) node[right] {~~~};
   \draw[->] (0,-1.2) -- (0,1.2) node[above] {$k_y$};
   \draw[very thick,cyan] (0,0) circle (1.0);
   \draw[very thick,blue] (0,0) circle (0.6600);
 \draw[very thick,->] (0,0) -- ( 1.0, 0.0) node[above=5pt,right=-1pt] {$z_1$};
 \draw[very thick,->] (0,0) -- ( 0.5, 0.8660) node[above=5pt,right=-1pt] {$z_2$};
 \draw[very thick,->] (0,0) -- (-0.5, 0.8660); 
 \draw[very thick,->] (0,0) -- (-1.0, 0.0   );
 \draw[very thick,->] (0,0) -- (-0.5,-0.8660); 
 \draw[very thick,->] (0,0) -- ( 0.5,-0.8660) node[above=-5pt,right=-1pt] {$z_3$};;
 \draw[very thick,->] (0.5, 0.8660) -- ( 1.0, 0.0);
\end{tikzpicture}%
\begin{tikzpicture}[scale=1.30,>=stealth]
   \draw[->] (-1.3,0) -- (1.3,0) node[right] {$k_x$};
   \draw[->] (0,-1.2) -- (0,1.2) node[above] {$k_y$};
   \draw[very thick,cyan] (0,0) circle (1.0);
   \draw[very thick,blue] (0,0) circle (0.6600);
 \draw[very thick,->] (0,0) -- (  0.6600,  0.0000) node[above=4pt,right=-3.0pt] {$w_1$};
 \draw[very thick,->] (0,0) -- (  0.3300,  0.5716) node[above=3pt,right=-2.5pt] {$w_2$};
 \draw[very thick,->] (0,0) -- ( -0.3300,  0.5716); 
 \draw[very thick,->] (0,0) -- ( -0.6600,  0.0000);
 \draw[very thick,->] (0,0) -- ( -0.3300, -0.5716); 
 \draw[very thick,->] (0,0) -- (  0.3300, -0.5716) node[above=-4pt,right=-3.5pt] {$w_3$};
 \draw[very thick,->] (0.3300,  0.5716) -- (  0.6600,  0.0000);
\end{tikzpicture}%
\end{minipage}
\hfill~~~
\begin{minipage}{0.30\linewidth}
\begin{equation*}
\begin{split}
\dot{z}_1 &=  \Qzh z_2 z_3 \\
\dot{w}_1 &=  \Qwh w_2 w_3 \phantom{\qquad\qquad\qquad}
\end{split}
\end{equation*}
\end{minipage}
\hfill}

%
%
\hbox to \hsize{%
\begin{minipage}{0.60\linewidth}
\begin{tikzpicture}[scale=1.30,>=stealth]
   \draw[->] (-1.3,0) -- (1.3,0) node[right] {~~~};
   \draw[->] (0,-1.2) -- (0,1.2) node[above] {$k_y$};
   \draw[very thick,cyan] (0,0) circle (1.0);
   \draw[very thick,blue] (0,0) circle (0.6600);
 \draw[very thick,->] (0,0) -- (   1.0000,  0.0000) node[above=5pt,right=-1pt] {$z_1$};
 \draw[very thick,->] (0,0) -- (   0.7822,  0.6230) node[above=4pt,right=-1pt] {$z_4$};
 \draw[very thick,->] (0,0) -- (  -0.7822,  0.6230);
 \draw[very thick,->] (0,0) -- (  -1.0000,  0.0000);
 \draw[very thick,->] (0,0) -- (  -0.7822, -0.6230);
 \draw[very thick,->] (0,0) -- (   0.7822, -0.6230) node[below=5pt,right=-1pt] {$z_5$};
 \draw[very thick,->] (0,0) -- (   0.2178,  0.6230) node[above=3pt,right=-3pt] {$w_5$};
 \draw[very thick,->] (0,0) -- (  -0.2178,  0.6230);
 \draw[very thick,->] (0,0) -- (  -0.2178, -0.6230);
 \draw[very thick,->] (0,0) -- (   0.2178, -0.6230) node[below=4pt,right=-3pt] {$w_4$};
 \draw[very thick,->] (   0.7822,   0.6230) -- ( 1, 0);
 \draw[very thick,->] (   0.7822,  -0.6230) -- ( 1, 0);
\end{tikzpicture}%
\begin{tikzpicture}[scale=1.30,>=stealth]
   \draw[->] (-1.3,0) -- (1.3,0) node[right] {$k_x$};
   \draw[->] (0,-1.2) -- (0,1.2) node[above] {$k_y$};
   \draw[very thick,cyan] (0,0) circle (1.0);
   \draw[very thick,blue] (0,0) circle (0.6600);
 \draw[very thick,->] (0,0) -- (  0.3300,  0.9440) node[above=2pt,right=0pt] {$z_6$};
 \draw[very thick,->] (0,0) -- ( -0.3300,  0.9440);
 \draw[very thick,->] (0,0) -- ( -0.3300, -0.9440);
 \draw[very thick,->] (0,0) -- (  0.3300, -0.9440) node[above=-6pt,right=0pt] {$z_7$};
 \draw[very thick,->] ( 0.3300,  0.9440) -- (  0.6600,  0.0000);
 \draw[very thick,->] (0,0) -- (  0.6600,  0.0000) node[above=4pt,right=-3.0pt] {$w_1$};
 \draw[very thick,->] (0,0) -- ( -0.6600,  0.0000);
   \draw (0.3,0) arc (0:70.73:0.3);
   \draw (0.3,0) arc (0:-70.73:0.3);
   \draw (0.3,0) arc (0:35:0.3) -- (0.95,0.63);
   \draw (1.0,0.63) node[above=0pt, right=-3pt] {$\theta_z$};
\end{tikzpicture}%
\end{minipage}
\hfill~~~
\begin{minipage}{0.30\linewidth}
\begin{equation*}
\begin{split}
\dot{z}_1 &= \Qzw(z_4 w_4 + z_5 w_5) \phantom{\qquad\qquad\qquad} \\
\dot{w}_1 &= \Qzz z_6 z_7 
\end{split}
\end{equation*}
\end{minipage}
\hfill}

%
\hbox to \hsize{%
\begin{minipage}{0.60\linewidth}
\begin{tikzpicture}[scale=1.30,>=stealth]
   \draw[->] (-1.3,0) -- (1.3,0) node[right] {~~~};
   \draw[->] (0,-1.2) -- (0,1.2) node[above] {$k_y$};
   \draw[very thick,cyan] (0,0) circle (1.0);
   \draw[very thick,blue] (0,0) circle (0.6600);
 \draw[very thick,->] (0,0) -- (   1.0000,  0.0000) node[above=5pt,right=-1pt] {$z_1$};
 \draw[very thick,->] (0,0) -- (  -1.0000,  0.0000);
 \draw[very thick,->] (0,0) -- (  0.5000,  0.4308) node[above=3pt,right=-4pt] {$w_6$};
 \draw[very thick,->] (0,0) -- ( -0.5000,  0.4308);
 \draw[very thick,->] (0,0) -- ( -0.5000, -0.4308);
 \draw[very thick,->] (0,0) -- (  0.5000, -0.4308) node[below=4pt,right=-4pt] {$w_7$};
 \draw[very thick,->] (  0.5000,  0.4308) -- (1,0);
 \draw (0.2500,-0.2154) arc (-40.75:40.75:0.33);
 \draw (0.315261, -0.0975217) -- (1.14640, -0.354624)
            node[below=6pt, right=-4pt] {$\theta_w$};
\end{tikzpicture}%
\begin{tikzpicture}[scale=1.30,>=stealth]
   \draw[->] (-1.3,0) -- (1.3,0) node[right] {$k_x$};
   \draw[->] (0,-1.2) -- (0,1.2) node[above] {$k_y$};
   \draw[very thick,cyan] (0,0) circle (1.0);
   \draw[very thick,blue] (0,0) circle (0.6600);
 \draw[very thick,->] (0,0) -- (  0.7576,  0.6527) node[above=2pt,right=0pt] {$z_8$};
 \draw[very thick,->] (0,0) -- ( -0.7576,  0.6527);
 \draw[very thick,->] (0,0) -- ( -0.7576, -0.6527);
 \draw[very thick,->] (0,0) -- (  0.7576, -0.6527) node[above=-6pt,right=0pt] {$z_9$};
 \draw[very thick,->] (  0.7576,  0.6527) -- (0.66,0);
 \draw[very thick,->] (  0.7576, -0.6527) -- (0.66,0);
 \draw[very thick,->] (0,0) -- (  0.6600,  0.0000) node[above=4pt,right=-3.0pt] {$w_1$};
 \draw[very thick,->] (0,0) -- (  0.0976,  0.6527);
 \draw[very thick,->] (0,0) -- ( -0.0976,  0.6527) node[above=4pt,left=-5.0pt] {$w_9$};
 \draw[very thick,->] (0,0) -- ( -0.6600,  0.0000);
 \draw[very thick,->] (0,0) -- ( -0.0976, -0.6527) node[below=5pt,left=-5.0pt] {$w_8$};
 \draw[very thick,->] (0,0) -- (  0.0976, -0.6527);
\end{tikzpicture}%
\end{minipage}
\hfill~~~
\begin{minipage}{0.30\linewidth}
\begin{equation*}
\begin{split}
\dot{z}_1 &=  \Qww w_6 w_7 \\
\dot{w}_1 &=  \Qwz (w_8z_8 + w_9z_9) \phantom{\qquad\qquad\qquad}
\end{split}
\end{equation*}
\end{minipage}
\hfill}

\caption{Three-wave interactions with two wavenumbers $k=1$ (outer circle) and
$k=q$ (inner circle) that influence the evolution of $z_1$ (left column) and
$w_1$ (centre column). Each vector is labelled with the amplitude ($z_1$, 
$w_1$, \dots) of the corresponding mode.
 First row: three wave vectors of the same length (three long or three short).
 Middle row: two long wave vectors and one short, defining an angle $\theta_z=2\arccos(q/2)$.
 Bottom row: one long wave vector and two short, defining an angle $\theta_w=2\arccos(1/2q)$.
 This last case only occurs when $q>\frac{1}{2}$.
 In all cases, the right column gives the quadratic terms in the amplitude
 equation that result from the three-wave interactions depicted to the left.
 }

\label{fig:TWI}

\end{figure}


\section{Nonlinear three-wave interactions}

\label{sec:3WI}

We first consider patterns in the variations of a real scalar field~$U(x,y,t)$. Assume
the system forms patterns with two distinct length scales. More specifically, and without
loss of generality, we assume that waves with wavenumbers $k=1$ and $k=q$ ($q<1$) become
unstable and have growth rates $r_1$ and $r_q$ respectively. At onset, the pattern
$U(x,y,t)$ will contain a combination of Fourier modes $e^{i\bk\cdot\bx}$, with $|\bk|=q$
or $|\bk|=1$. We write, close to onset,
 \begin{equation}
 U=  \sum_{\bq_j} w_j(t) e^{i\bq_j\cdot\bx} +
     \sum_{\bk_j} z_j(t) e^{i\bk_j\cdot\bx} +
     \text{higher order terms},
 \label{eq:leadingorderU}
 \end{equation}
where $\bq_j$ are wavevectors on the circle $|\bk|=q$, with mode
amplitudes~$w_j(t)$, and
      $\bk_j$ are wavevectors on the circle $|\bk|=1$, with mode
amplitudes~$z_j(t)$. The overall pattern $U$ is real, so waves come in equal and opposite pairs with complex conjugate amplitudes. 

The time evolution of the complex mode amplitudes is
influenced by nonlinear combinations of other mode amplitudes. The particular
combinations that arise are determined by {\red the lengths and orientations of 
the wavevectors}, in a manner that can be explained by focusing on one mode on
each circle and examining the lowest-order combinations that influence the
chosen mode.

The two modes we choose are $z_1(t) e^{i\bk_1\cdot\bx}$ and $w_1(t) e^{i\bq_1\cdot\bx}$,
as well as their complex conjugates, illustrated in Figure~\ref{fig:TWI}. We will develop
an ordinary differential equation (ODE) for each mode amplitude and express it as a
truncated Taylor series. The linear terms in the evolution equation for $z_1$ and $w_1$
are simply $r_1z_1$ and $r_qw_1$, respectively, and the starting point for the ODEs
describing the evolution of each mode amplitude is
 \begin{equation}
 \begin{split}
 {\dot z}_1 &= r_1 z_1 + \mbox{nonlinear terms,} \\
 {\dot w}_1 &= r_q w_1 + \mbox{nonlinear terms.}
 \end{split}
 \label{eq:linearODEs}
 \end{equation}
Nonlinear functions of~$U$, as written in~(\ref{eq:leadingorderU}), will involve products
of modes and therefore sums of wavevectors. The combinations of modes that
influence~$z_1$ and~$w_1$ will be those whose wavevectors add up to~$\bk_1$ and~$\bq_1$
respectively. The lowest-order nonlinear terms are quadratic, arising when two vectors
(of length~$1$ or~$q$) add up to $\bk_1$ or~$\bq_1$. The simplest interactions involve
modes at~$60^\circ$. The wave vectors in these so-called hexagonal states can be arranged
in an equilateral triangle; see Figure~\ref{fig:TWI}, top row. If $\bk_1=\bk_2+\bk_3$ (all
of length~$1$), and $\bq_1=\bq_2+\bq_3$ (all of length~$q$) then the equations for ${\dot
z}_1$ and ${\dot w}_1$ will have the terms $\Qzh z_2z_3$ and $\Qwh w_2w_3$, where $z_2$,
$z_3$, $w_2$ and $w_3$ are the amplitudes of modes with wavevectors $\bk_2$, $\bk_3$,
$\bq_2$ and $\bq_3$ respectively, and $\Qzh$ and $\Qwh$ are coefficients.

As well as equilateral triangles, one may have isosceles triangles with one short and two
long sides (Figure~\ref{fig:TWI}, middle row) and triangles with one long and two short
sides (Figure~\ref{fig:TWI}, bottom row). The latter case can only happen if
$q>\frac{1}{2}$.
The two isosceles triangles define related angles
 \begin{equation}
 \theta_z=2\arccos(q/2), \qquad
 \theta_w=2\arccos(1/2q),
 \label{eq:thetazw}
 \end{equation}
as seen in Figure~\ref{fig:TWI}~\cite{RucSilSke2012}.
These triangles lead, in different combinations, to contributions
indicated in the right column of Figure~\ref{fig:TWI}, where $\Qzw$, $\Qzz$, $\Qww$ and
$\Qwz$ are further coefficients. The mode amplitudes are numbered in order of appearance 
in Figure~\ref{fig:TWI}.
The end result is that, at quadratic order, there are 8~modes that couple to each of
$z_1$ and~$w_1$:
 \begin{equation}
 \begin{split}
 {\dot z}_1 &= \cdots + \Qzh z_2 z_3
             + \Qzw (z_4 w_4 + z_5 w_5) + \Qww w_6 w_7 + \cdots\\
 {\dot w}_1 &= \cdots + \Qwh w_2 w_3
             + \Qzz z_6 z_7 + \Qwz (w_8z_8 + w_9z_9) + \cdots
 \end{split}
 \label{eq:quadraticODEs}
 \end{equation}
These 16 additional modes, 8~with wavenumber~1 and 8~with wavenumber~$q$, will
each couple to {\red up to} 8~further modes, and each of these further modes will couple
to {\red up to} 8~more, as so on, as outlined in~\cite{RucSilSke2012}.


\begin{figure}

\hbox to \hsize{%
\begin{tikzpicture}[scale=1.30,>=stealth]
   \draw[->] (-1.3,0) -- (1.3,0) node[right] {$k_x$};
   \draw[->] (0,-1.2) -- (0,1.2) node[above] {$k_y$};
   \draw[very thick,cyan] (0,0) circle (1.0);
   \draw[very thick,blue] (0,0) circle (0.37796);
   \draw[very thick,->] (0,0) --  (    -0.188982,    -0.981981);
   \draw[very thick,->] (0,0) --  (     0.188982,    -0.981981);
   \draw[very thick,->] (0,0) --  (    -0.755929,    -0.654654);
   \draw[very thick,->] (0,0) --  (     0.755929,    -0.654654);
   \draw[very thick,->] (0,0) --  (    -0.944911,    -0.327327);
   \draw[very thick,->] (0,0) --  (    -0.188982,    -0.327327);
   \draw[very thick,->] (0,0) --  (     0.188982,    -0.327327);
   \draw[very thick,->] (0,0) --  (     0.944911,    -0.327327);
   \draw[very thick,->] (0,0) --  (    -0.377964,      0.00000);
   \draw[very thick,->] (0,0) --  (     0.377964,      0.00000);
   \draw[very thick,->] (0,0) --  (    -0.944911,     0.327327);
   \draw[very thick,->] (0,0) --  (    -0.188982,     0.327327);
   \draw[very thick,->] (0,0) --  (     0.188982,     0.327327);
   \draw[very thick,->] (0,0) --  (     0.944911,     0.327327);
   \draw[very thick,->] (0,0) --  (    -0.755929,     0.654654);
   \draw[very thick,->] (0,0) --  (     0.755929,     0.654654);
   \draw[very thick,->] (0,0) --  (    -0.188982,     0.981981);
   \draw[very thick,->] (0,0) --  (     0.188982,     0.981981);
 \draw[very thick,->] (0.188982,     0.981981) -- (0.377964,      0.00000);
 \draw (0.0, -1.15) node[below] {\strut(a) $q=1/\sqrt{7}$};
\end{tikzpicture}%
\hfill
\begin{tikzpicture}[scale=1.30,>=stealth]
   \draw[->] (-1.3,0) -- (1.3,0) node[right] {$k_x$};
   \draw[->] (0,-1.2) -- (0,1.2) node[above] {$k_y$};
   \draw[very thick,cyan] (0,0) circle (1.0);
   \draw[very thick,blue] (0,0) circle (0.5176);
%
 \draw[very thick,->] (0,0) -- (  0.9659,  0.2588);
 \draw[very thick,->] (0,0) -- (  0.7071,  0.7071);
 \draw[very thick,->] (0,0) -- (  0.2588,  0.9659);
 \draw[very thick,->] (0,0) -- ( -0.2588,  0.9659);
 \draw[very thick,->] (0,0) -- ( -0.7071,  0.7071);
 \draw[very thick,->] (0,0) -- ( -0.9659,  0.2588);
 \draw[very thick,->] (0,0) -- ( -0.9659, -0.2588);
 \draw[very thick,->] (0,0) -- ( -0.7071, -0.7071);
 \draw[very thick,->] (0,0) -- ( -0.2588, -0.9659);
 \draw[very thick,->] (0,0) -- (  0.2588, -0.9659);
 \draw[very thick,->] (0,0) -- (  0.7071, -0.7071);
 \draw[very thick,->] (0,0) -- (  0.9659, -0.2588);
 \draw[very thick,->] (0,0) -- (  0.5176,  0.0000);
 \draw[very thick,->] (0,0) -- (  0.4483,  0.2588);
 \draw[very thick,->] (0,0) -- (  0.2588,  0.4483);
 \draw[very thick,->] (0,0) -- ( -0.0000,  0.5176);
 \draw[very thick,->] (0,0) -- ( -0.2588,  0.4483);
 \draw[very thick,->] (0,0) -- ( -0.4483,  0.2588);
 \draw[very thick,->] (0,0) -- ( -0.5176, -0.0000);
 \draw[very thick,->] (0,0) -- ( -0.4483, -0.2588);
 \draw[very thick,->] (0,0) -- ( -0.2588, -0.4483);
 \draw[very thick,->] (0,0) -- (  0.0000, -0.5176);
 \draw[very thick,->] (0,0) -- (  0.2588, -0.4483);
 \draw[very thick,->] (0,0) -- (  0.4483, -0.2588);
 \draw[very thick,->] (  0.5176,  0.0000)  -- (  0.9659,  0.2588);
 \draw[very thick,->] (  0.2588,  0.9659)  -- (  0.5176,  0.0000);
 \draw (0.0, -1.15) node[below] {\strut(b) $q=\sqrt{2-\sqrt{3}}$};
 \draw                    (  0.3300, -0.9440) node[above=-8pt,right=0pt] {\phantom{\smash{$\bk_2$}}};
\end{tikzpicture}
\hfill
\begin{tikzpicture}[scale=1.30,>=stealth]
   \draw[->] (-1.3,0) -- (1.3,0) node[right] {$k_x$};
   \draw[->] (0,-1.2) -- (0,1.2) node[above] {$k_y$};
   \draw[very thick,cyan] (0,0) circle (1.0);
   \draw[very thick,blue] (0,0) circle (0.66);
%
 \draw[->] (0,0) -- (  0.3300,  0.9440);
 \draw[->] (0,0) -- ( -0.3300,  0.9440);
 \draw[->] (0,0) -- ( -0.3300, -0.9440);
 \draw[->] (0,0) -- (  0.3300, -0.9440);
 \draw[->] (0,0) -- (  0.6600,  0.0000);
 \draw[->] (0,0) -- ( -0.6600,  0.0000);
 \draw[->] (0,0) -- (  0.9993,  0.0386);
 \draw[->] (0,0) -- (  0.7576,  0.6527);
 \draw[->] (0,0) -- (  0.3300,  0.9440);
 \draw[->] (0,0) -- (  0.1859,  0.9826);
 \draw[->] (0,0) -- ( -0.1859,  0.9826);
 \draw[->] (0,0) -- ( -0.3300,  0.9440);
 \draw[->] (0,0) -- ( -0.7576,  0.6527);
 \draw[->] (0,0) -- ( -0.9993,  0.0386);
 \draw[->] (0,0) -- ( -0.9993, -0.0386);
 \draw[->] (0,0) -- ( -0.7576, -0.6527);
 \draw[->] (0,0) -- ( -0.3300, -0.9440);
 \draw[->] (0,0) -- ( -0.1859, -0.9826);
 \draw[->] (0,0) -- (  0.1859, -0.9826);
 \draw[->] (0,0) -- (  0.3300, -0.9440);
 \draw[->] (0,0) -- (  0.7576, -0.6527);
 \draw[->] (0,0) -- (  0.9993, -0.0386);
 \draw[->] (0,0) -- (  0.6600,  0.0000);
 \draw[->] (0,0) -- (  0.5717,  0.3298);
 \draw[->] (0,0) -- (  0.2417,  0.6142);
 \draw[->] (0,0) -- ( -0.2417,  0.6142);
 \draw[->] (0,0) -- ( -0.5717,  0.3298);
 \draw[->] (0,0) -- ( -0.6600,  0.0000);
 \draw[->] (0,0) -- ( -0.5717, -0.3298);
 \draw[->] (0,0) -- ( -0.2417, -0.6142);
 \draw[->] (0,0) -- (  0.2417, -0.6142);
 \draw[->] (0,0) -- (  0.5717, -0.3298);
 \draw[->] (0,0) -- (  0.9993,  0.0386);
 \draw[->] (0,0) -- (  0.9824,  0.1868);
 \draw[->] (0,0) -- (  0.8848,  0.4659);
 \draw[->] (0,0) -- (  0.8463,  0.5328);
 \draw[->] (0,0) -- (  0.7576,  0.6527);
 \draw[->] (0,0) -- (  0.6520,  0.7582);
 \draw[->] (0,0) -- (  0.4018,  0.9157);
 \draw[->] (0,0) -- (  0.3300,  0.9440);
 \draw[->] (0,0) -- (  0.1859,  0.9826);
 \draw[->] (0,0) -- ( -0.1859,  0.9826);
 \draw[->] (0,0) -- ( -0.3300,  0.9440);
 \draw[->] (0,0) -- ( -0.4018,  0.9157);
 \draw[->] (0,0) -- ( -0.6520,  0.7582);
 \draw[->] (0,0) -- ( -0.7576,  0.6527);
 \draw[->] (0,0) -- ( -0.8463,  0.5328);
 \draw[->] (0,0) -- ( -0.8848,  0.4659);
 \draw[->] (0,0) -- ( -0.9824,  0.1868);
 \draw[->] (0,0) -- ( -0.9993,  0.0386);
 \draw[->] (0,0) -- ( -0.9993, -0.0386);
 \draw[->] (0,0) -- ( -0.9824, -0.1868);
 \draw[->] (0,0) -- ( -0.8848, -0.4659);
 \draw[->] (0,0) -- ( -0.8463, -0.5328);
 \draw[->] (0,0) -- ( -0.7576, -0.6527);
 \draw[->] (0,0) -- ( -0.6520, -0.7582);
 \draw[->] (0,0) -- ( -0.4018, -0.9157);
 \draw[->] (0,0) -- ( -0.3300, -0.9440);
 \draw[->] (0,0) -- ( -0.1859, -0.9826);
 \draw[->] (0,0) -- (  0.1859, -0.9826);
 \draw[->] (0,0) -- (  0.3300, -0.9440);
 \draw[->] (0,0) -- (  0.4018, -0.9157);
 \draw[->] (0,0) -- (  0.6520, -0.7582);
 \draw[->] (0,0) -- (  0.7576, -0.6527);
 \draw[->] (0,0) -- (  0.8463, -0.5328);
 \draw[->] (0,0) -- (  0.8848, -0.4659);
 \draw[->] (0,0) -- (  0.9824, -0.1868);
 \draw[->] (0,0) -- (  0.9993, -0.0386);
 \draw[->] (0,0) -- (  0.6600,  0.0000);
 \draw[->] (0,0) -- (  0.5717,  0.3298);
 \draw[->] (0,0) -- (  0.5163,  0.4112);
 \draw[->] (0,0) -- (  0.4830,  0.4498);
 \draw[->] (0,0) -- (  0.3304,  0.5714);
 \draw[->] (0,0) -- (  0.2417,  0.6142);
 \draw[->] (0,0) -- (  0.0976,  0.6527);
 \draw[->] (0,0) -- ( -0.0976,  0.6527);
 \draw[->] (0,0) -- ( -0.2417,  0.6142);
 \draw[->] (0,0) -- ( -0.3304,  0.5714);
 \draw[->] (0,0) -- ( -0.4830,  0.4498);
 \draw[->] (0,0) -- ( -0.5163,  0.4112);
 \draw[->] (0,0) -- ( -0.5717,  0.3298);
 \draw[->] (0,0) -- ( -0.6600,  0.0000);
 \draw[->] (0,0) -- ( -0.5717, -0.3298);
 \draw[->] (0,0) -- ( -0.5163, -0.4112);
 \draw[->] (0,0) -- ( -0.4830, -0.4498);
 \draw[->] (0,0) -- ( -0.3304, -0.5714);
 \draw[->] (0,0) -- ( -0.2417, -0.6142);
 \draw[->] (0,0) -- ( -0.0976, -0.6527);
 \draw[->] (0,0) -- (  0.0976, -0.6527);
 \draw[->] (0,0) -- (  0.2417, -0.6142);
 \draw[->] (0,0) -- (  0.3304, -0.5714);
 \draw[->] (0,0) -- (  0.4830, -0.4498);
 \draw[->] (0,0) -- (  0.5163, -0.4112);
 \draw[->] (0,0) -- (  0.5717, -0.3298);
 \draw (0.0, -1.15) node[below] {\strut(c) $q=0.66$};
\end{tikzpicture}

\hfill}

\caption{Pattern wavevectors involved in three-wave interactions for different values 
 of~$q=|\bq_j|$ from~(\ref{eq:leadingorderU}). 
 (a)~$q=1/\sqrt{7}=0.3780$ ($\theta_z=158.2^\circ$), 
 (b)~$q=\sqrt{2-\sqrt{3}}=0.5176$ ($\theta_z=150^\circ$, $\theta_w=30^\circ$), 
 (c)~$q=0.66$ ($\theta_z=141.5^\circ$, $\theta_w=81.5^\circ$).
 The angles $\theta_z$ and $\theta_w$ are defined in~(\ref{eq:thetazw}) and 
 Figure~\ref{fig:TWI}.}

\label{fig:threeexamples}

\end{figure}
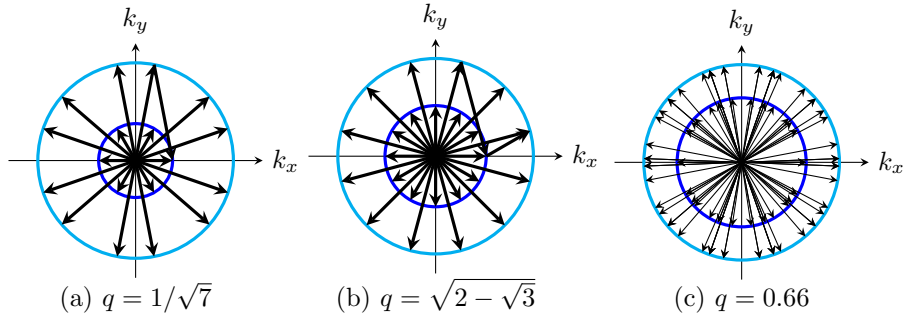

One might ask, ``where does it all end?'' The answer depends on~$q$, as explained 
in~\cite{RucSilSke2012}. For
$q<\frac{1}{2}$, two short vectors added together do not extend to the outer
circle, and so the interactions in Figure~\ref{fig:TWI} (bottom row) do not
exist, and the end result is, for example, six modes on the inner circle and
12~on the outer, as in Figure~\ref{fig:threeexamples}(a). This case can lead to
superlattice patterns~\cite{KudPieGol1998,EpsFin2006,TopSil2002}
or quasipatterns~\cite{IooRuc2020}. For
$q=2\sin\left(\frac{\pi}{12}\right)=\sqrt{2-\sqrt{3}}=0.5176$, the angles $\theta_z$ and $\theta_w$ (defined in Figure~\ref{fig:TWI}) are
$150^\circ$ and $30^\circ$ respectively, and so all possible three-wave
interactions can be accommodated within a set of 12~vectors of length~1
interleaved with 12~vectors of length~$q$, as in
Figure~\ref{fig:threeexamples}(b). This special value of~$q$ is the only one
in the range $\frac{1}{2}<q<1$ where three-wave interactions generate a finite 
number of modes~\cite{RucSilSke2012}. For all other $\frac{1}{2}<q<1$,
an infinite number of modes is generated, as illustrated in
Figure~\ref{fig:threeexamples}(c) for a generic choice of~$q$.

Of course, equations~(\ref{eq:linearODEs}) and~(\ref{eq:quadraticODEs}) go only
up to quadratic order. At cubic order, every wave on the two circles couples to
every other wave, since $\bk_1=\bk_1+\bk_j-\bk_j=\bk_1+\bq_j-\bq_j$, for any
vectors~$\bk_j$ and~$\bq_j$. Given a finite set of modes as in
Figure~\ref{fig:threeexamples}(a,b), one can work out the amplitude equations,
calculate quadratic and cubic (and higher if needed) coefficients, and analyse
which solutions are possible and stable. Doing this for complex periodic
patterns is challenging because dozens of modes are involved. For
quasipatterns, there is the additional complication that this process, where
the small-amplitude pattern is expressed as a power series in a small
parameter, leads to divergent series~\cite{RucRuc2003,IooRuc2010},
though existence of quasipatterns has been proved in the Swift--Hohenberg 
equation~\cite{BraIooSto2017} and in Rayleigh--B\'enard convection~\cite{BraIoo2019}. 
The case of a
potentially infinite set of modes (Figure~\ref{fig:threeexamples}c) is
challenging. 

The purpose of our present work is to see how far considerations from
just the quadratic level can help understand the outcome when both circles in
Fourier space are (potentially) fully occupied. The example we use to
illustrate the ideas is the two-layer Brusselator model, in 
Section~\ref{sec:Brusselator}. Before discussing this model, we review what is
known about the role of three-wave interactions in the formation of complex 
spatiotemporal patterns and outline our hypotheses regarding how
quadratic coefficients would influence observed patterns.


\section{Role of the quadratic coefficients}

\label{sec:quadraticcoefficients}

Figure~\ref{fig:threeexamples} gives the three qualitatively different possible
cases. If $q<\frac{1}{2}$, all three-wave interactions can be accommodated
within a set of 6~vectors of length~$q$ and 12 vectors of length~1, as in
Figure~\ref{fig:threeexamples}(a). This leads to 9~coupled complex amplitude
equations. 
The resulting patterns are spatially periodic when
$\cos\theta_z$ and $\sqrt{3}\sin\theta_z$ are both rational, which
happens for a dense but measure zero set of~$q$. Otherwise, the
resulting patterns are quasiperiodic~\cite{IooRuc2020}.
The second case, as in Figure~\ref{fig:threeexamples}(b), with $q=\sqrt{2-\sqrt{3}}=0.5176$, leads to 12~vectors of length~$q$ and 12 vectors of length~1, necessitating 12~complex amplitude equations. In these two cases, with a finite number of amplitude equations (which will be explored in more detail elsewhere), standard nonlinear methods can be employed to obtain
equilibrium points, and (to some extent) their stability, bifurcations, and so forth. 
In the third case, with $\frac{1}{2}<q\neq\sqrt{2-\sqrt{3}}<1$ as in 
Figure~\ref{fig:threeexamples}(c), three-wave interactions lead to coupling 
between an infinite number of modes, and so there is the possibility of an infinite number of 
amplitude equations. In this scenario, it is not clear that 
standard nonlinear methods will yield useful information.

All three cases involve sets of interacting waves, with the strongest interactions happening
between groups of three. 
We illustrate a single set of three interacting waves by taking two outer vectors coupling to an inner one, 
with wavevectors
$\bk_6+\bk_7=\bq_1$ and amplitudes $z_6$, $z_7$,~$w_1$, as in 
Figure~\ref{fig:TWI}
(middle row, middle column). The amplitude 
equations in this case are of the form:
 \begin{equation}
 \label{eq:cubic}
 \begin{split}
 \dot{z}_6 &= r_1 z_6 + \Qzw \bar{z}_7 w_1 + (A_{z} |z_6|^2 + A_{zz}|z_7|^2 + A_{zw}|w_1|^2)z_6\\ 
 \dot{z}_7 &= r_1 z_7 + \Qzw \bar{z}_6 w_1 + (A_{zz}|z_6|^2 + A_{z} |z_7|^2 + A_{zw}|w_1|^2)z_7\\ 
 \dot{w}_1 &= r_q w_1 + \Qzz z_6 z_7  + (A_{wz}|z_6|^2 + A_{wz}|z_7|^2 + A_{w}|w_1|^2)w_1.
 \end{split}
 \end{equation}
Here, $A_{z}$, $A_{zz}$, $A_{zw}$, $A_{w}$ and $A_{wz}$ are cubic coefficients that
depend on the details of the problem and that can in principle be calculated
from governing equations.

Porter and Silber~\cite{PorSil2004} investigated (\ref{eq:cubic}) in detail and
found that the dynamics depends on the product of quadratic coefficients~$\Qzw\Qzz$,
as well as the linear and cubic coefficients.
Typically, when $\Qzw\Qzz$ is positive, there are stable equilibria
and no time-dependent states. On the other hand, when $\Qzw\Qzz$ is
negative, in addition to stable equilibria, time-periodic solutions and chaotic
solutions are possible via Hopf and global bifurcations. In the positive case,
the $z$ and $w$ modes can act to reinforce each other, while in the negative
case, there can be time-dependent competition between $z$ and~$w$ modes. The
same conclusion applies equally to the three-wave interaction between two~$w$
and one~$z$ mode (Figure~\ref{fig:TWI}, bottom row, left column). Here, the
relevant combination of quadratic coefficients is $\Qwz\Qww$.

These considerations led Rucklidge \etal~\cite{RucSilSke2012} to hypothesise 
how the combinations of quadratic coefficients would influence patterns, 
essentially supposing that the qualitative conclusion of~\cite{PorSil2004} 
applies also when there are many sets of interacting waves, even though
each individual wave has three-wave interactions with several combinations
of modes, as discussed above.
Steady patterns should be expected when $\Qzw\Qzz>0$ and $\Qwz\Qww>0$, 
and
time-dependent patterns should be possible when one or both pairs of
quadratic coefficients are of opposite sign. When $q>\frac{1}{2}$, 
complex patterns, with modes at many different orientations, as in 
Figure~\ref{fig:threeexamples}(c), may be possible. 

Before developing these ideas further, for the purposes of this paper, we
distinguish between different types of patterns. \emph{Simple} patterns
are stripes, hexagons {\red (or symmetry-broken hexagons)} of either critical wavelength.
There are also \emph{rhombs}, here taken to 
mean patterns with two modes of equal amplitude on one circle coupled to a third
mode on the other circle.
\emph{Superlattice}
patterns are dominated by 12~modes at one wavenumber and 6~at the other; 
here we blur the distinction between spatially periodic superlattice patterns
and quasipatterns~\cite{IooRuc2020}.
With $q<\frac{1}{2}$, there is only one type of superlattice pattern, while 
with $\frac{1}{2}<q<1$, there are two types, with six modes on one circle and
twelve on the other, either way around.
\emph{Regular twelve-fold quasipatterns} have 12 modes (equally spaced) at each wavenumber, as
illustrated in Figure~\ref{fig:threeexamples}(b). 
The patterns discussed so far may have \emph{defects}, 
which can evolve over long timescales~\cite{CroHoh1993}.
\emph{Complex} patterns have
large numbers of modes, at both wavenumbers, coupled through three-wave
interactions, as illustrated in Figure~\ref{fig:threeexamples}(c), but are
not simple patterns with defects as defined here.
Time
dependent (periodic, chaotic) versions of each of these types of patterns are
also possible, evolving over shorter timescales. 
We reserve the term \emph{spatiotemporal chaos} for the case
when complex patterns have persistent chaotic dynamics with many positive Lyapunov exponents as 
in~\cite{PauEinFis2007}.

With this classification in mind, we extend the hypotheses
of~\cite{RucSilSke2012} as detailed in the points below, and as summarised in
Table~\ref{table:predictions}. Here by \emph{finding
a pattern}, we mean that there are combinations of $r_1$ and $r_q$ where that
pattern is an asymptotic state obtained when starting from random initial
conditions in a domain large enough to accommodate a wide range of
wavevector orientations on the two critical circles. 

\begin{itemize}

\item
In all cases, we expect to find steady simple patterns such as stripes and
hexagons, possibly with {\red broken symmstry or with} defects.

\item
In addition, with $q<\frac{1}{2}$, we expect to find steady superlattice
patterns with wavevectors as in Figure~\ref{fig:threeexamples}(a). We may also find rhombs. If
$\Qzw\Qzz$ is negative, we expect to find time-dependent superlattice patterns
(and rhombs) with the same wavevectors, and also spatiotemporal chaos, with all
wavevectors on the two circles being active. We do not expect to find steady
complex patterns.

\item
With $q>\frac{1}{2}$, we expect to find both types of steady superlattice
patterns. We also expect to find steady complex patterns, with large numbers of
wavevectors on both circles. The combinations of quadratic coefficients
relevant to the two superlattice cases are $\Qzw\Qzz$ and $\Qwz\Qww$
respectively: if the relevant combination is negative, we expect to find
time-dependent superlattice patterns with the same wavevectors. If either or
both combination is negative, we expect to find spatiotemporal chaos,
with all wavevectors on the two circles being active. If $\Qzw\Qzz$ and
$\Qwz\Qzz$ are both negative, we expect to find time dependence more readily.
In general, we expect to find spatiotemporal chaos more readily than in the
$q<\frac{1}{2}$ case.

\item
For the special value $q=\sqrt{2-\sqrt{3}}=0.5176$, we expect to find steady
twelve-fold quasipatterns with wavevectors as in
Figure~\ref{fig:threeexamples}(b). If one or both of $\Qzw\Qzz$ or $\Qwz\Qww$
is negative, we expect to find time-dependent quasipatterns and
spatiotemporal chaos.

\end{itemize}

\begin{table}
\begin{tabular*}{\linewidth}{c|c|c}
\hline        
$q$ & 
\begin{minipage}[l]{0.40\linewidth}
$\Qzw\Qzz>0$ and $\Qwz\Qww>0$ 
\strut\end{minipage}
&  
\begin{minipage}{0.40\linewidth}\strut
$\Qzw\Qzz<0$ or $\Qwz\Qww<0$
\strut\end{minipage}\\
\hline
$q<\frac{1}{2}$ &
 \begin{minipage}[l]{0.40\linewidth}\strut
 Steady superlattice patterns
 (only $\Qzw\Qzz$ is relevant)
 \strut\end{minipage}
& 
 \begin{minipage}[l]{0.40\linewidth}\strut
 Steady and oscillatory superlattice patterns,
 possibly spatiotemporal chaos
 (only $\Qzw\Qzz$ is relevant)
 \strut\end{minipage}
\\
\hline 
$q>\frac{1}{2}$
&
 \begin{minipage}[l]{0.40\linewidth}\strut
 Steady superlattice patterns of both types
 and steady complex patterns
 \strut\end{minipage}
& 
 \begin{minipage}[l]{0.40\linewidth}\strut
 Steady and time-dependent superlattice patterns of both types,
 steady complex patterns and spatiotemporal chaos
 \strut\end{minipage}
\\
\hline
\begin{minipage}[l]{0.12\linewidth}\strut
$\sqrt{2-\sqrt{3}}$
$=0.5176$
\end{minipage}
&
 \begin{minipage}[l]{0.40\linewidth}\strut
 Steady twelve-fold quasipatterns
 \strut\end{minipage}
& 
 \begin{minipage}[l]{0.40\linewidth}\strut
 Steady and time-dependent twelve-fold quasipatterns,
 steady complex patterns and spatiotemporal chaos
 \strut\end{minipage}
\\
\hline
\end{tabular*}
\caption{Patterns that \emph{a priori} we expect to find in different circumstances, 
in addition to steady simple patterns (stripes and hexagons).}
 \label{table:predictions}
\end{table}

These considerations neglect the roles that the hexagonal 
quadratic coefficients $\Qzh$ and~$\Qwh$ might play.


\section{Two-layer Brusselator model}

\label{sec:Brusselator}

The Brusselator~\cite{PriLef1968,LavPosRom2009} is a canonical model of a reaction--diffusion system. More specifically, it describes an autocatalytic chemical reaction,
\begin{equation}
\begin{split}
A \to X \\
2X + Y \to 3X \\
B + X \to Y + D \\
X \to E.
\end{split}
\end{equation}
The products $D,E$ are generally not of interest because they do not enter into the autocatalysis. Therefore, we restrict attention to the reactants $X,Y,A,B$. In the Brusselator, it is assumed that $A,B$ are present in great excess, and thus can be treated as constants. Allowing for spatial diffusion, and using the standard theories of reaction kinetics, we write down rate laws for $X,Y$ as the differential equations
\begin{equation}
\label{eq:basicbrusselator}
\begin{split}
\frac{\partial X} {\partial t} &= A + X^2 Y - BX - X +D_X \nabla^2 X,\\
\frac{\partial Y} {\partial t} &= BX - X^2 Y+D_Y \nabla^2 Y.
\end{split}
\end{equation}
Through abuse of notation, we have now let $A,B,X,Y \geq 0$ represent concentrations of these chemicals rather than symbolising the chemicals themselves. Here, $X,Y$ are time and space dependent chemical concentrations and, as assumed, $A,B$ are constant.

Eq.~(\ref{eq:basicbrusselator}) has a spatially homogeneous steady state solution, namely $X = A$, $Y = B/A$. We adopt shifted coordinates for the dependent variables, letting $X = A + U$, $Y = B/A + V$ so that the equilibrium becomes the trivial one, $U = 0$, $V=0$. In these coordinates, (\ref{eq:basicbrusselator}) is
 \begin{equation}
 \begin{split}
 \frac{\partial U}{\partial t} &=
 (B-1)U + A^2V + D_U\nabla^2U + \frac{B}{A}U^2 + 2AUV + U^2V,\\
 \frac{\partial V}{\partial t} &=
 -BU - A^2V + D_V\nabla^2V - \frac{B}{A}U^2 - 2AUV - U^2V.
 \end{split}
 \label{eq:Brusselator1}
 \end{equation}
The chemical concentrations are $U(\bx,t)$ and $V(\bx,t)$, where $\bx$ is the planar spatial coordinate $\bx = (x,y)$. The diffusion constants have been relabelled for clarity of notation, that is, $D_U = D_X$ and $D_V = D_Y$.

As in~\cite{YanDolZha2002,CatMcNTop2012}, we consider a two-layer Brusselator model. The 
layers are coupled together ``diffusively,'' manifesting as linear terms with coefficients~$\alpha,\beta \geq 0$:
 \begin{equation}
 \begin{split}
 \frac{\partial U_1}{\partial t} &=
 (B-1)U_1 + A^2V_1 + \DUi\nabla^2U_1 + {\alpha(U_2-U_1)} + \hbox{NLT}(U_1,V_1),\\
 \frac{\partial V_1}{\partial t} &=
 -BU_1 - A^2V_1 + \DVi\nabla^2V_1 + {\beta(V_2-V_1)} - \hbox{NLT}(U_1,V_1),\\
 \frac{\partial U_2}{\partial t} &=
 (B-1)U_2 + A^2V_2 + \DUj\nabla^2U_2 + {\alpha(U_1-U_2)} + \hbox{NLT}(U_2,V_2),\\
 \frac{\partial V_2}{\partial t} &=
 -BU_2 - A^2V_2 + \DVj\nabla^2V_2 + {\beta(V_1-V_2)} - \hbox{NLT}(U_2,V_2).
 \end{split}
 \label{eq:Brusselator2}
 \end{equation}
Here, $U_{1,2}(\bx,t)$ and $V_{1,2}(\bx,t)$ are chemical concentrations in each layer. For convenience, we have used shorthand to represent the nonlinear terms,
\begin{equation}
\hbox{NLT}(U,V) \equiv \frac{B}{A}U^2 + 2AUV + U^2V.
\end{equation}
We have assumed that $A$ and $B$ do not vary across layers, meaning that each excess reactant is present in the same amount in each layer. For all calculations in the remainder of this paper, we take $A=3$ and $B=9$ as our standard parameter values, as chosen in~\cite{YanDolZha2002} {\red to model the CIMA reaction}.

\section{Linear theory}

\label{sec:linear}


If we drop the nonlinear terms in~(\ref{eq:Brusselator2}), we can solve the resulting
linear PDE in terms of modes $e^{i\bk\cdot\bx}$ that grow as
$e^{\sigma t}$. Here, $\sigma$ is a growth rate that depends on the
wavenumber~$k=|\bk|$. The linear problem is represented by a $4\times4$
Jacobian matrix~$J$,
 \begin{equation}
 \bJ = 
 \begin{pmatrix}
 B-1 - \DUi k^2 \mathrlap{{}-\alpha} & A^2                      & \alpha                  & 0                        \\
 -B                      & - A^2 - \DVi k^2 \mathrlap{{}-\beta} & 0                       & \beta                    \\
 \alpha                  & 0                        & B-1 - \DUj k^2 \mathrlap{{}-\alpha} & A^2                      \\
 0                       & \beta                    & -B                      & -A^2 - \DVj k^2 - \beta
 \end{pmatrix}.
 \label{eq:Jacobian2}
 \end{equation}
The growth rates~$\sigma$ are the eigenvalues of~$\bJ$ and satisfy
the characteristic equation
 \begin{equation}
 \sigma^4 + C_3 \sigma^3 +
            C_2 \sigma^2 +
            C_1 \sigma +
            C_0 = 0,
 \label{eq:characteristic}
 \end{equation}
where the coefficients $C_0$, \dots, $C_3$ are (cumbersome) polynomial functions of nine 
parameters: $A$,
$B$, $\DUi$, $\DVi$, $\DUj$, $\DVj$, $\alpha$, $\beta$ and the wavenumber~$k$.

If we were to choose $\DUi=\DUj$ and $\DVi=\DVj$, as done in~\cite{CatMcNTop2012},
then the $4\times4$ matrix~$\bJ$ can be decomposed in to two $2\times2$ parts.
However, in this case, it turns out that two of the quadratic coefficients
vanish in the weakly nonlinear theory. To avoid this degeneracy, we take an
alternative approach, allowing the diffusion constants to be different in the
two layers. As mentioned in Section~\ref{sec:Brusselator}, the experimental context is
that we take the chemistry to be identical in the two layers (meaning $A,B$ do not depend
on layer) but take the substrates to be different, so their diffusion properties will be
different.

Rather than fix the values of the parameters, we are aiming to explore the range of
outcomes close to the codimension-two point where patterns with two length scales are
simultaneously unstable, for a range of values of the wavenumber ratio. Therefore, we
seek parameter values for which~$\sigma$, when viewed as a function of~$k$, takes on
certain values at local maxima. For example, if $\sigma$ has a local maximum at $k=1$ and
$k=q$, for some choice of~$q$, then four conditions must be satisfied: $\sigma=r_1$ at
$k=1$, $\sigma=r_q$ at $k=q$ and $\frac{d\sigma}{dk}=0$ at $k=1, q$. The
derivative~$\frac{d\sigma}{dk}$ can be obtained from~(\ref{eq:characteristic}) by
differentiating with respect to~$k$:
 \begin{equation}
 \left(4\sigma^3 + 3C_3\sigma^2 + 2C_2\sigma + C_1\right)\frac{d\sigma}{dk} +
            \frac{dC_3}{dk} \sigma^3 +
            \frac{dC_2}{dk} \sigma^2 +
            \frac{dC_1}{dk} \sigma   +
            \frac{dC_0}{dk}          = 0,
 \label{eq:dcharacteristic}
 \end{equation}
The four conditions result in four equations for the nine parameters listed above (but
with $k$ replaced by~$q$), with two additional parameters $\sigma(1)=r_1$ and
$\sigma(q)=r_q$. This means that seven of the parameters can be specified, and four found 
by solving the equations. For example, we take
our base parameter values $A=3$ and $B=9$~\cite{YanDolZha2002} and choose
$q=\sqrt{2-\sqrt{3}}=0.5176$, appropriate for twelve-fold
quasipatterns~\cite{Mul1994,LifPet1997}. Additionally, we choose $r_1=r_q=0$ in order
to be at the codimension-two point, and we choose
$\alpha=\beta=1$. Recall that $\alpha$ and~$\beta$ control the diffusion of the two
chemicals between the two layers, while $\DUi$, $\DUj$, $\DVi$ and~$\DVj$ control the
diffusion of the chemicals within each layer. For this choice of seven parameters, the
resulting four polynomial equations for the four remaining unknowns ($\DUi$, $\DUj$,
$\DVi$ and~$\DVj$) can be worked out; the simplest (shortest) of these is
 \begin{equation}
 \begin{split}
 & \DUi\DUj\DVi\DVj + 10\DUi\DUj\DVi + 10\DUi\DUj\DVj \\
              & \qquad  {} -  7\DUi\DVi\DVj -  7\DUj\DVi\DVj
 + 99 \DUi\DUj + 48 \DVi\DVj \\
 & \qquad {} + 11 \DUi\DVi + 11 \DUj\DVj - 70 \DUi\DVj - 70 \DUj\DVi \\
 & \qquad {} + 117 \DUi + 117 \DUj - 87 \DVi - 87 \DVj + 135 = 0.
 \end{split}
 \end{equation}
The coefficients in this (and the other three equations) depend on the choice
that we made for $A$, $B$, $\alpha$, $\beta$, $q$, $r_1$ and~$r_q$.

\begin{figure}
 \hbox to \hsize{\hfill%
  \includegraphics[width=0.90\hsize]{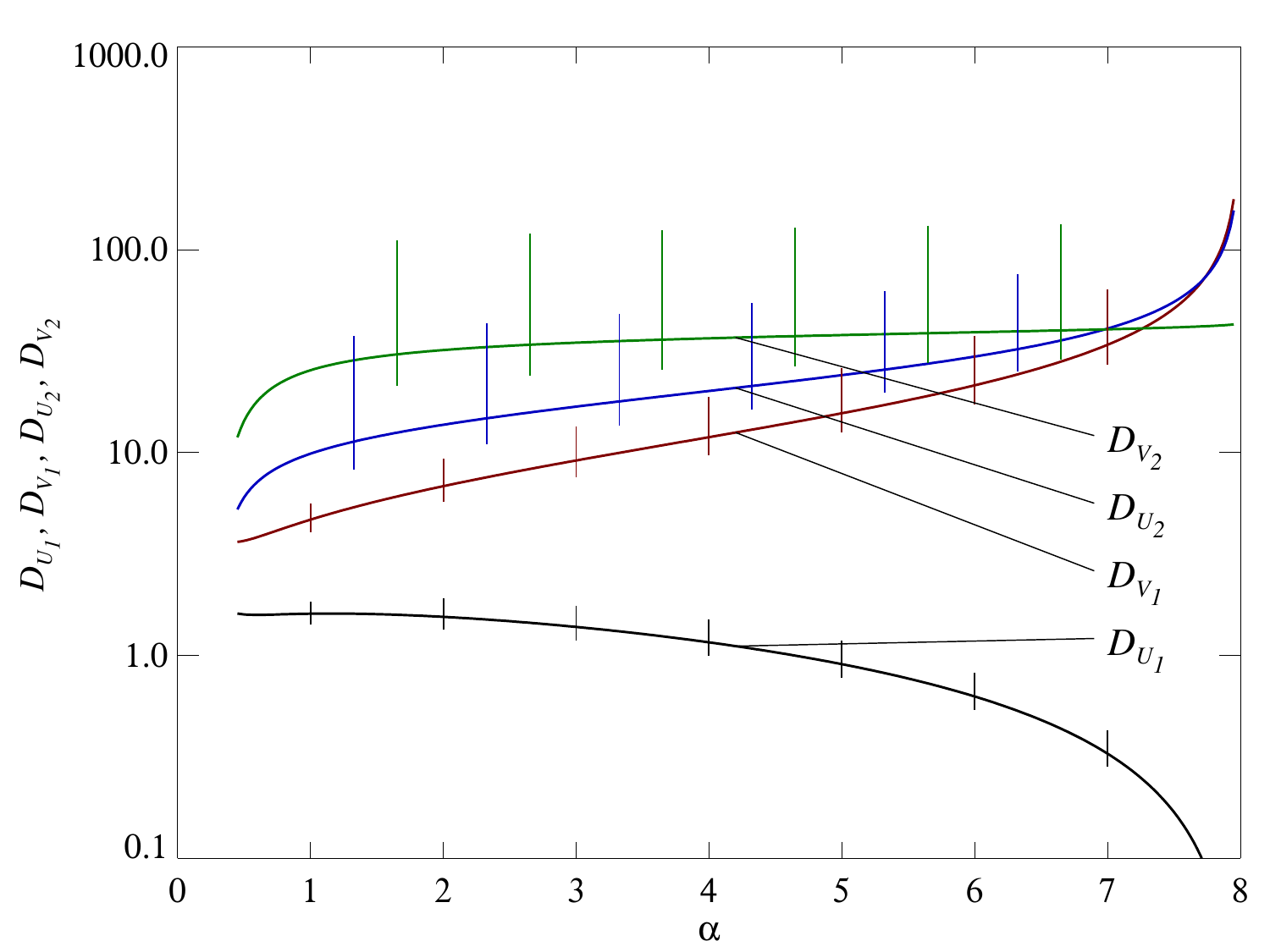}\hfill}
\caption{Linear theory for the two-layer Brusselator model, for 
$q=\sqrt{2-\sqrt{3}}=0.5176$, $\beta=1$,
$A=3$, $B=9$, $r_1=0$ and $r_q=0$. We plot {\red $\DUi$ (black), 
$\DVi$ (red), $\DUj$ (blue) and $\DVj$ (green)}
for $0.45\leq\alpha\leq7.95$.
The vertical lines indicate how the diffusion coefficients vary
as $q$ ranges from $q=0.25$ to $q=0.66$: these coefficients generally decrease as $q$~increases.
{\red The vertical lines are at integer values of~$\alpha$ for $\DUi$ and $\DVi$,
and are shifted by $\frac{1}{3}$ for~$\DUj$ and
                by $\frac{2}{3}$ for~$\DVj$.}
Sample numerical values of the diffusion constants are given in 
Table~\ref{table:wnlt} and in full in~\cite{DataRepository}.}
 \label{fig:linear}
 \end{figure}

We solve the four polynomial equations numerically, using
\textsc{Bertini}~\cite{BatHauSom2013}. For our current choice of parameters, there are 24
solutions, of which eight~are real but only two (related by relabelling the two layers)
are real and positive:
 \begin{equation}
 \DUi = 1.6046, \quad
 \DVi = 4.6663, \quad
 \DUj = 9.8682, \quad
 \DVj = 25.448.
 \label{eq:example_diffusion}
 \end{equation}
The number of real positive solutions varies with $\alpha$ and~$\beta$: for
example, with $\beta=1$ and $q=\sqrt{2-\sqrt{3}}$, there are none for
$\alpha\leq0.31$ or $\alpha\geq8$, and two (related by relabelling) for
$0.32\leq\alpha\leq7.99$. We plot the four diffusion coefficients as functions
of~$\alpha$ for $\beta=1$ in Figure~\ref{fig:linear} in black for
$q=\sqrt{2-\sqrt{3}}=0.5176$, 
{\red with vertical lines indicating how the diffusion coefficients vary with~$q$,}
keeping $A$, $B$, $r_1$ and $r_q$ fixed. 
Sample numerical values of the diffusion constants 
for different choices of~$\alpha$ and~$q$ are given in Table~\ref{table:wnlt} (below) 
and in full in~\cite{DataRepository}.

We observe from Figure~\ref{fig:linear} that the range from the smallest to the
largest values of the diffusion constants appears to diverge as $\alpha$
approaches~$8$. {\red The same happens for other choices of~$q$.} In addition, the ordering of the diffusion constants changes
around $\alpha=7$: for $\alpha<7$, we have $\DUi<\DVi$ and $\DUj<\DVj$, which
seems experimentally reasonable, in that one chemical diffuses slower than the
other in either substrate, while for $\alpha>7$, this is not true. For later
calculations, we will choose $\alpha$ to vary between~$1$ and~$7$. The
experimental relevance of the larger values of~$\alpha$ should be treated with
caution. 

\begin{figure}
 \hbox to \hsize{\hfill%
  \includegraphics[width=0.90\hsize]{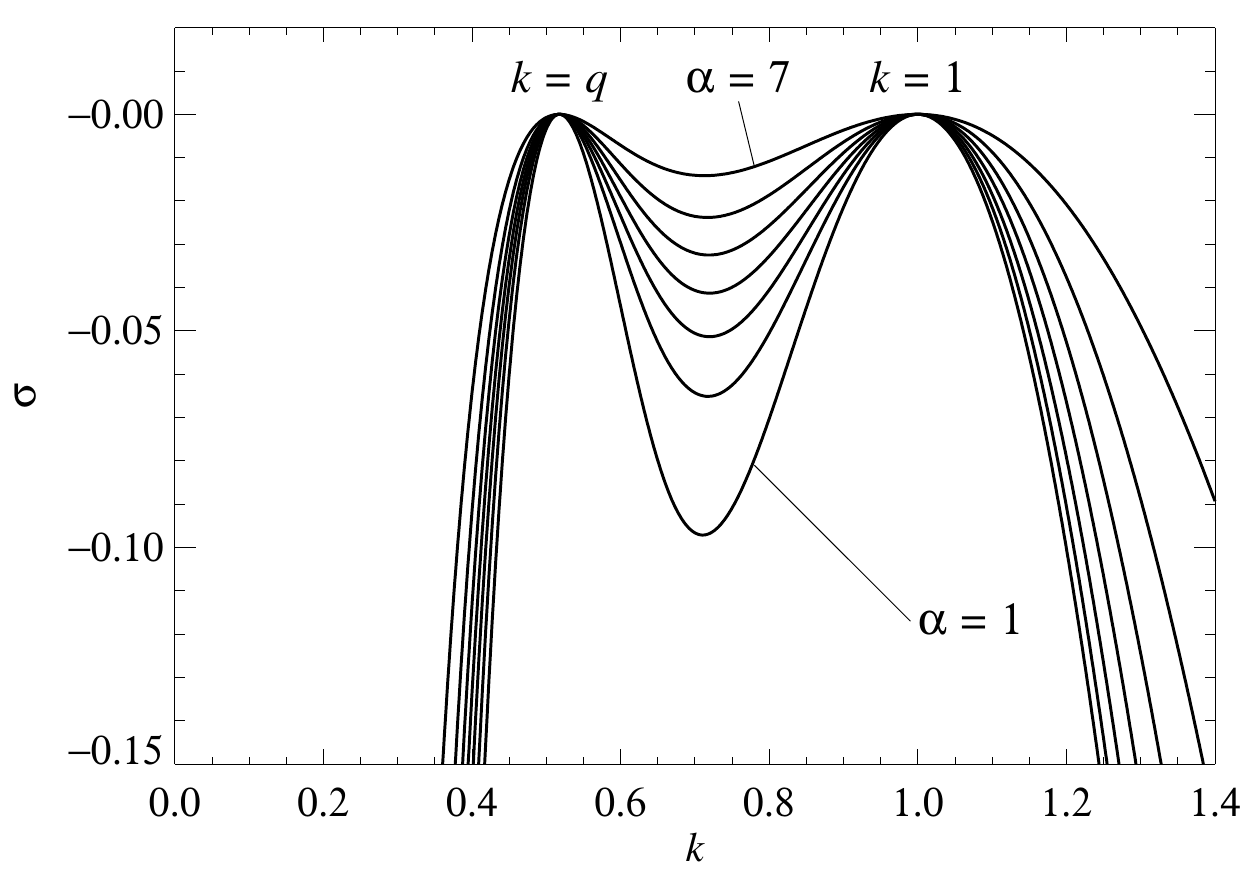}\hfill}
\caption{Dispersion relation: the largest eigenvalue~$\sigma(k)$ plotted as a function of 
wavenumber~$k$, for $q=\sqrt{2-\sqrt{3}}=0.5176$, $\beta=1$, and 
$\alpha=1,2,\dots,7$.}
 \label{fig:dispersion}
 \end{figure}

We conclude our discussion of the linear theory with a sample dispersion
relation ($\sigma(k)$ plotted as a function of wavenumber~$k$) in
Figure~\ref{fig:dispersion}, for $q=\sqrt{2-\sqrt{3}}$ and for a range 
of~$\alpha$, at the codimension-two point $r_1=r_q=0$.
The eigenvalue is maximal at $k=q$ and $k=1$, but the minimum
for $q<k<1$ is about $-0.1$ for $\alpha=1$, and only about $-0.015$ for
$\alpha=7$. 

In this dispersion relation, we control the separation and heights of the
growth-rate maxima by varying $q$, $r_1$ and~$r_1$, and solving for the four diffusion 
coefficients. 
For smaller~$q$, the peaks are well separated and reasonably sharp, while for
larger~$q$ the peaks are closer together, broader and the depth of the minimum is less. For this reason, we limit ourselves
to $q\leq0.66$. Since we want to keep an interval of negative growth-rate
between $k=q$ and $k=1$, we (mostly) limit ourselves to $r_1\leq0.01$
and $r_q\leq0.01$.

\section{Weakly nonlinear theory}

\label{sec:wnlt}

Once the uniform state $U_1=V_1=U_2=V_2=0$ becomes linearly unstable, solutions will grow
exponentially until nonlinear effects become important. The first of these are
the three-wave interactions. The weakly nonlinear theory is 
standard~\cite{VerDeWDew1992,PenPer2001,JudSil2000}, though made more complicated here
because of the codimension-two bifurcation and because there are four scalar fields in
(\ref{eq:Brusselator2}). We are concerned only with the leading order effect of
three-wave interactions, and so we need to compute only up to second order in the weakly
nonlinear theory.

We write
 \begin{equation}
 \bu = \begin{pmatrix} U_1 \\ V_1 \\ U_2 \\ V_2 \end{pmatrix},
 \end{equation}
and write the PDE~(\ref{eq:Brusselator2}) as
 \begin{equation}
 \frac{d\bu}{dt} = \cL\bu + \mbox{\bNLT$(\bu)$},
 \label{eq:abstractPDE}
 \end{equation}
where $\cL$ is a linear operator representing the linear terms, and all the
nonlinear terms in~(\ref{eq:Brusselator2}) are in $\mbox{\bNLT$(\bu)$}$.
To explore the properties of solutions close to~$\bu=0$,
we introduce a small parameter~$\epsilon\ll1$, and we expand $\bu$ in powers
of~$\epsilon$:
 \begin{equation}
 \bu = \epsilon \bu_1 + \epsilon^2 \bu_2 + \cdots
 \end{equation}
Recall that in Section~\ref{sec:linear}, we computed values of $\DUi$, $\DVi$,
$\DUj$ and $\DVj$ such that the linear operator~$\cL$ had zero eigenvalues ($r_1=r_q=0$) at
two wavenumbers, $k=1$ and $k=q$, at given values of $A$, $B$, $\alpha$ and
$\beta$. We now suppose that the linear operator is perturbed by an
order~$\epsilon$ amount so that the growth rates~$r_1$ at $k=1$ and $r_q$ at
$k=q$ are order~$\epsilon$. In practice we perturb $\DUi$, $\DVi$, $\DUj$ and
$\DVj$ and do it in such a way that there are local maxima in the growth rate
remain at $k=1$ and $k=q$. We can scale $r_1\rightarrow\epsilon r_1$
and $r_q\rightarrow\epsilon r_q$ and write the linear operator~$\cL$ as
 \begin{equation}
 \cL = \cL_0 + \epsilon \cL_1 + \cdots,
 \end{equation}
where $\cL_0$ is a singular linear operator, and $\epsilon\cL_1$ is the largest
part of the perturbation of the linear operator from~$\cL_0$. Finally, we scale
time so that $d/dt\rightarrow\epsilon d/dt$. With these choices of scaling, 
the time derivative, the linear terms, and the lowest-order nonlinear terms all 
appear at the same order.
Substituting into~(\ref{eq:abstractPDE}), we have
  \begin{equation}
 \epsilon^2\frac{d\bu_1}{dt} = \epsilon\cL_0\bu_1 +
                               \epsilon^2\cL_0\bu_2 +
                               \epsilon^2\cL_1\bu_1 +
\epsilon^2\mbox{\bNLT$_2(\bu_1)$} + \cO(\epsilon^3),
 \label{eq:expandedPDE}
 \end{equation}
where $\mbox{\bNLT$_2$}$ represents the quadratic nonlinear terms.

The operator~$\cL_0$ is singular: $\cL_0 e^{i\bk\cdot\bx}\bv_1=0$ whenever
$|\bk|=1$, and $\cL_0 e^{i\bq\cdot\bx}\bv_q=0$ whenever $|\bq|=q$, where
$\bv_1$ and $\bv_q$ are the eigenvectors of the zero eigenvalues of
the Jacobian matrix~(\ref{eq:Jacobian2}), with~$k$ replaced by~$1$ and~$q$
respectively. We normalise the eigenvectors so that $\bv_1\cdot\bv_1=1$ and
$\bv_q\cdot\bv_q=1$.
Following the example of Section~\ref{sec:linear} and~(\ref{eq:example_diffusion}), 
with $A=3$, $B=9$, $\alpha=1$ and
$\beta=1$, we find
 \begin{equation}
 \bv_1 = \begin{pmatrix}\phm0.8416\\ -0.5198\\ \phm0.1377\\ -0.0496\end{pmatrix}
 \qquad\text{and}\qquad
 \bv_q = \begin{pmatrix}\phm0.5288\\ -0.4550\\ \phm0.6201\\ -0.3589\end{pmatrix}.
 \label{eq:example_eigenvectors}
 \end{equation}
With these eigenvectors, the general solution to $\cL_0\bu_1=0$ is similar to the
expression in~(\ref{eq:leadingorderU}):
 \begin{equation}
 \bu_1 =  \left(\sum_{\bq_j} w_j(t) e^{i\bq_j\cdot\bx}\right) \bv_q +
          \left(\sum_{\bk_j} z_j(t) e^{i\bk_j\cdot\bx}\right) \bv_1,
 \label{eq:leadingorderu}
 \end{equation}
where 
$\{\bq_j\}$ and $\{\bk_j\}$ are arbitrary sets of vectors on the two circles
$|\bq_j|=q$ and $|\bk_j|=1$. Writing $\bu_1$ in this way solves the 
$\cO(\epsilon)$ part of~(\ref{eq:expandedPDE}). 


The $\cO(\epsilon^2)$ part of~(\ref{eq:expandedPDE}) is
 \begin{equation}
 \frac{d\bu_1}{dt} =           \cL_0\bu_2 +
                               \cL_1\bu_1 + \mbox{\bNLT$_2(\bu_1)$}.
 \label{eq:expandedPDE2}
 \end{equation}
Recall that~$\cL_0$ is singular and so cannot simply be inverted to
find~$\bu_2$ as a function of~$\bu_1$. Thus, before solving for~$\bu_2$, a
solvability condition must be imposed. The standard method is to
define an inner product between vector-valued functions $\bff(\bx)$ and
$\bg(\bx)$ on the domain~$\Omega$ of the problem:
 \begin{equation}
 \big\langle \bff,\bg\big\rangle = 
   \frac{1}{|\Omega|}\int_{\Omega} {\bar \bff}(\bx) \cdot \bg(\bx) \, d\bx,
 \label{eq:innerproduct}
 \end{equation}
where ${\bar \bff}$ is the complex conjugate of~$\bff$ and 
$|\Omega|$~is the area of the domain. 
We define~$\cLdag_0$, 
the adjoint of~$\cL_0$, by requiring that
 \begin{equation}
 \big\langle \bff,\cL_0 \bg\big\rangle = \big\langle\cLdag_0 \bff,\bg\big\rangle 
 \label{eq:adjoint}
 \end{equation}
for all $\bff$ and~$\bg$. We restrict to
functions on~$\Omega$ that satisfy periodic boundary conditions. In this case,
the adjoint operator~$\cLdag_0$ is just the transpose of~$\cL_0$. Having
defined~$\cLdag_0$, we solve $\cLdag_0 e^{i\bk\cdot\bx}\bvdag_1=0$ and
$\cLdag_0 e^{i\bq\cdot\bx}\bvdag_q=0$ to find the normalised adjoint
eigenvectors~$\bvdag_1$ and~$\bvdag_q$, with $|\bk|=1$ and $|\bq|=q$.
For our example, these are
 \begin{equation}
 \bvdag_1 = \begin{pmatrix}0.8416\\ 0.5198\\ 0.1377\\ 0.0496\end{pmatrix}
 \qquad\text{and}\qquad
 \bvdag_q = \begin{pmatrix}0.5288\\ 0.4550\\ 0.6201\\ 0.3589\end{pmatrix}.
 \label{eq:example_adjoinnt_eigenvectors}
 \end{equation}
Then, for any~$\bu_2$,
 \begin{equation}
 \begin{split}
 \Big\langle e^{i\bk_1\cdot\bx}\bvdag_1,\cL_0\bu_2\Big\rangle &= 
 \Big\langle \cLdag_0e^{i\bk_1\cdot\bx}\bvdag_1,\bu_2\Big\rangle = 0,\\
 \Big\langle e^{i\bq_1\cdot\bx}\bvdag_q,\cL_0\bu_2\Big\rangle &= 
 \Big\langle \cLdag_0e^{i\bq_1\cdot\bx}\bvdag_q,\bu_2\Big\rangle = 0,
 \end{split}
 \end{equation}
where $\bk_1$ and $\bq_1$ represent any
vectors on the two critical circles.
Thus, taking the inner products of $e^{i\bk_1\cdot\bx}\bvdag_1$ and 
$e^{i\bq_1\cdot\bx}\bvdag_q$ with~(\ref{eq:expandedPDE2})
results in 
the solvability conditions
 \begin{equation}
 \begin{split}
 \Big\langle e^{i\bk_1\cdot\bx}\bvdag_1,
                \frac{d\bu_1}{dt}
                \Big\rangle &= 
 \Big\langle e^{i\bk_1\cdot\bx}\bvdag_1,
                \cL_1\bu_1 + \mbox{\bNLT$_2(\bu_1)$}
                \Big\rangle,\\
 \Big\langle e^{i\bq_1\cdot\bx}\bvdag_q,
                \frac{d\bu_1}{dt}
                \Big\rangle &= 
 \Big\langle e^{i\bq_1\cdot\bx}\bvdag_q,
                \cL_1\bu_1 + \mbox{\bNLT$_2(\bu_1)$}
                \Big\rangle.
 \end{split}
 \label{eq:solvability2}
 \end{equation}
Taking $\bu_1$ to be made up of waves with wavevectors from all the
combinations of wavevectors in Figure~\ref{fig:TWI} results in amplitude
equations (including the values of the coefficients) up to quadratic order, as
written in~(\ref{eq:quadraticODEs}). We will take two specific examples,
focusing only on the quadratic coefficients, and compute $\Qzh$, $\Qzz$
and~$\Qzw$. For these, we need $\mbox{\bNLT$_2(\bu_1)$}$, which is (for $A=3$ and
$B=9$):
 \begin{equation}
 \mbox{\bNLT$_2(\bu_1)$} = \begin{pmatrix}
                         \phm 3 U_1^2 + 6 U_1 V_1 \\
                            - 3 U_1^2 - 6 U_1 V_1 \\
                         \phm 3 U_2^2 + 6 U_2 V_2 \\
                            - 3 U_2^2 - 6 U_2 V_2
                         \end{pmatrix},
 \end{equation}
where $(U_1,V_1,U_2,V_2)$ are the four entries in~$\bu_1$.

To calculate the various quadratic coefficients described in 
Section~\ref{sec:3WI}, we take the combinations of wavevectors
appropriate for each coefficient. 
For~$\Qzh$, we write
 \begin{equation}
 \bu_1 =  \left(z_1(t) e^{i\bk_1\cdot\bx} +
                z_2(t) e^{i\bk_2\cdot\bx} +
                z_3(t) e^{i\bk_3\cdot\bx} 
          \right) \bv_1 + \text{c.c.},
 \label{eq:leadingorderu_qzh}
 \end{equation}
where $\bk_1=\bk_2+\bk_3$ as in the  
top left panel of Figure~\ref{fig:TWI}, 
$\bv_1$ is the eigenvector as in~(\ref{eq:example_eigenvectors})
and \hbox{c.c.}
stands for the complex conjugate. In this case, we have
 \begin{equation}
 \begin{split}
 U_1^2 &= \left(z_1 e^{i\bk_1\cdot\bx} +
                z_2 e^{i\bk_2\cdot\bx} +
                z_3 e^{i\bk_3\cdot\bx} + \text{c.c.} \right)^2
          \times
          \left(\bv_1^{(1)}\right)^2, \\
       &= \left(\cdots + 2 z_2 z_3 e^{i\bk_1\cdot\bx} + \cdots\right)
          \times
          \left(\bv_1^{(1)}\right)^2, \\
U_1V_1 &= \left(z_1 e^{i\bk_1\cdot\bx} +
                z_2 e^{i\bk_2\cdot\bx} +
                z_3 e^{i\bk_3\cdot\bx} + \text{c.c.} \right)^2
          \times
          \left(\bv_1^{(1)}\bv_1^{(2)}\right), \\
       &= \left(\cdots + 2 z_2 z_3 e^{i\bk_1\cdot\bx} + \cdots\right)
          \times
          \left(\bv_1^{(1)}\bv_1^{(2)}\right),
 \end{split}
 \end{equation}
where we have highlighted the $e^{i\bk_1\cdot\bx}$ term, and $\bv_1^{(1)}$ and
$\bv_1^{(2)}$ are the first and second entries in the vector~$\bv_1$
in~(\ref{eq:example_eigenvectors}). There are similar expressions for $U_2^2$
and $U_2V_2$, involving $\bv_1^{(3)}$ and $\bv_1^{(4)}$. The inner product with
$e^{i\bk_1\cdot\bx}\bvdag_1$ in the first line of the solvability condition
in~(\ref{eq:solvability2}) picks out the $e^{i\bk_1\cdot\bx}$ component of
$\mbox{\bNLT$_2(\bu_1)$}$, so we are left with
 \begin{equation}
 \left(\bvdag_1\cdot\bv_1\right){\dot z_1} =
 \mbox{linear term} +
    \bvdag_1\cdot
      \begin{pmatrix} 
         \phm 3\times2\left(\bv_1^{(1)}\right)^2 + 6\times2\bv_1^{(1)}\bv_1^{(2)}\\
             -3\times2\left(\bv_1^{(1)}\right)^2 - 6\times2\bv_1^{(1)}\bv_1^{(2)}\\
         \phm 3\times2\left(\bv_1^{(3)}\right)^2 + 6\times2\bv_1^{(3)}\bv_1^{(4)}\\
             -3\times2\left(\bv_1^{(3)}\right)^2 - 6\times2\bv_1^{(3)}\bv_1^{(4)}
      \end{pmatrix}
      z_2 z_3.
 \end{equation}
We have used the fact that $\bvdag_1$ is real. Dividing by $\bvdag_1\cdot\bv_1$
and matching to~(\ref{eq:quadraticODEs}) results in an expression for~$\Qzh$.
For the example set of parameters, $\Qzh=-0.7018$. With $r_1$ defined to be the
growth rate (on the slow time scale) of the wavenumber~$|\bk|=1$ modes, the linear term
above is $r_1\left(\bvdag_1\cdot\bv_1\right)z_1$. 

Similar calculations but for different choices of wavevectors yield~$\Qwh$, $\Qzw$
and~$\Qzz$, and $\Qwz$ and~$\Qww$. We illustrate with the calculation for
$\Qzz$ and~$\Qzw$, and write
 \begin{equation}
 \bu_1 =  \left(z_6(t) e^{i\bk_6\cdot\bx} +
                z_7(t) e^{i\bk_7\cdot\bx}
          \right) \bv_1 + 
          \left(w_1(t) e^{i\bq_1\cdot\bx}
          \right) \bv_q + 
          \text{c.c.},
 \label{eq:leadingorderu_qzz}
 \end{equation}
where $\bq_1=\bk_6+\bk_7$ as in the  
middle row centre panel of Figure~\ref{fig:TWI}.
In this case, we need the $e^{i\bq_1\cdot\bx}$ and $e^{i\bk_6\cdot\bx}$
components of $U_1^2$ and $U_1V_1$:
 \begin{equation}
 \begin{split}
 U_1^2 &= \left(2 z_6 z_7 e^{i\bq_1\cdot\bx} + \cdots\right)
          \times
          \left(\bv_1^{(1)}\right)^2 + {} \\
       & \qquad
          \left(2 w_1 {\bar z}_7 e^{i\bk_6\cdot\bx} + \cdots\right)
          \times
          \left(\bv_1^{(1)}\bv_q^{(1)}\right), \\
U_1V_1 &= \left(2 z_6 z_7 e^{i\bq_1\cdot\bx} + \cdots\right)
          \times
          \left(\bv_1^{(1)}\bv_1^{(2)}\right) + {} \\
        & \qquad
          \left(w_1 {\bar z}_7 e^{i\bk_6\cdot\bx} + \cdots\right)
          \times
          \left(\bv_1^{(1)}\bv_q^{(2)} + \bv_1^{(2)}\bv_q^{(1)}\right),
 \end{split}
 \end{equation}
again with similar expressions for $U_2^2$ and $U_2V_2$.
The inner product with 
$e^{i\bk_6\cdot\bx}\bvdag_1$ in the first line of the solvability condition 
in~(\ref{eq:solvability2}) picks out the $\Qzw {\bar z}_7 w_1$ term,
while the inner product with 
$e^{i\bq_1\cdot\bx}\bvdag_q$ in the second line of the solvability condition
picks out the $\Qzz z_6z_7$ term. These result in equations for the two quadratic 
coefficients:
 \begin{equation}
 \begin{split}
 \left(\bvdag_1\cdot\bv_1\right) \Qzw &=
 \bvdag_1\cdot\begin{pmatrix}
              \phm 3\times2\bv_1^{(1)}\bv_q^{(1)}
                  +6\times\left(\bv_1^{(1)}\bv_q^{(2)} + \bv_1^{(2)}\bv_q^{(1)}\right)\\
                  -3\times2\bv_1^{(1)}\bv_q^{(1)}
                  -6\times\left(\bv_1^{(1)}\bv_q^{(2)} + \bv_1^{(2)}\bv_q^{(1)}\right)\\
              \phm 3\times2\bv_1^{(3)}\bv_q^{(3)}
                  +6\times\left(\bv_1^{(3)}\bv_q^{(4)} + \bv_1^{(4)}\bv_q^{(3)}\right)\\
                  -3\times2\bv_1^{(3)}\bv_q^{(3)}
                  -6\times\left(\bv_1^{(3)}\bv_q^{(4)} + \bv_1^{(4)}\bv_q^{(3)}\right)
              \end{pmatrix},
 \\
 \left(\bvdag_q\cdot\bv_q\right) \Qzz &=
 \bvdag_q\cdot\begin{pmatrix}
              \phm 3\times2\left(\bv_1^{(1)}\right)^2
                  +6\times2 \bv_1^{(1)}\bv_1^{(2)}\\
                  -3\times2\left(\bv_1^{(1)}\right)^2
                  -6\times2 \bv_1^{(1)}\bv_1^{(2)}\\
              \phm 3\times2\left(\bv_1^{(3)}\right)^2
                  +6\times2 \bv_1^{(3)}\bv_1^{(4)}\\
                  -3\times2\left(\bv_1^{(3)}\right)^2
                  -6\times2 \bv_1^{(3)}\bv_1^{(4)}
              \end{pmatrix}.
 \end{split}
 \end{equation}
For the example set of parameters, $\Qzw=-0.8974$ and $\Qzz=-0.1997$.
Similar calculations yield $\Qwh=-0.5610$, and 
$\Qwz=-0.2623$ and $\Qww=-0.9263$ (only available since $q>\frac{1}{2}$).

%



\begin{figure}
 \hbox to \hsize{\hfill%
  \includegraphics[width=0.90\hsize]{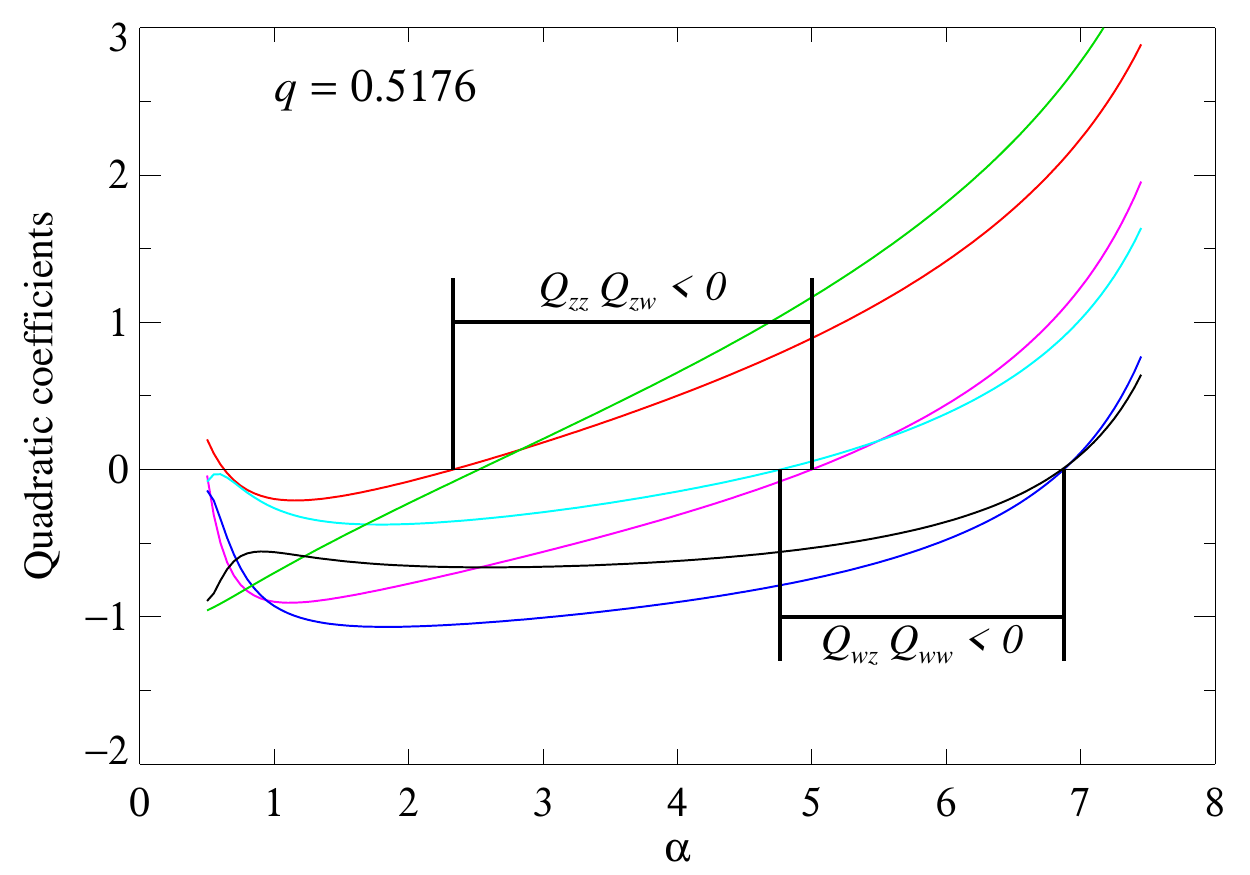}\hfill}
\caption{Weakly nonlinear theory for the two-layer Brusselator 
model~(\ref{eq:Brusselator2}), up to quadratic order. We take 
$\beta=1$ in~(\ref{eq:Brusselator2}) and 
$q=0.5176$ in~(\ref{eq:leadingorderU}). The six quadratic coefficients are 
 $\Qzh$~(green),
 $\Qwh$~(black),
 $\Qzz$~(red),
 $\Qzw$~(magenta),
 $\Qwz$~(cyan),
 $\Qww$~(blue).
In this case, the coefficients $\Qzw$ and $\Qzz$ have opposite
sign for $2.33<\alpha<5.00$, and $\Qwz$ and $\Qww$ have opposite
sign for $4.76<\alpha<6.87$. The values of diffusion coefficients are
as in Figure~\ref{fig:linear}. 
The data for this figure is available in full at~\cite{DataRepository}, with
selected values in Table~\ref{table:wnlt}.}
\label{fig:wnlt}
\end{figure}
            

\begin{table}\begingroup
\begin{small}
\setlength{\tabcolsep}{4.5pt} 
\begin{tabular*}{\linewidth}{c|c|cccc|cccccc}
\hline        
$q$ & $\alpha$ & $\DUi$ & $\DVi$ & $\DUj$ & $\DVj$ & $\Qzh$ & $\Qwh$ & $\Qzz$ & $\Qzw$ & $\Qwz$ & $\Qww$ \\
\hline                     
         &  $1$ & $1.75$ & $\phz5.2$ & $16.4$ & $44.0$ &    $-0.59$ &    $-0.52$ &    $-0.07$ &    $-0.85$ & --- & --- \\
         &  $2$ & $1.76$ & $\phz8.1$ & $21.2$ & $53.9$ & $\phm0.02$ &    $-0.57$ & $\phm0.04$ &    $-0.79$ & --- & --- \\
         &  $3$ & $1.59$ &    $11.2$ & $25.2$ & $57.9$ & $\phm0.57$ &    $-0.62$ & $\phm0.31$ &    $-0.60$ & --- & --- \\
$0.3780$ &  $4$ & $1.34$ &    $15.0$ & $29.5$ & $60.4$ & $\phm1.12$ &    $-0.64$ & $\phm0.64$ &    $-0.36$ & --- & --- \\
         &  $5$ & $1.05$ &    $20.0$ & $34.7$ & $62.3$ & $\phm1.71$ &    $-0.62$ & $\phm1.05$ &    $-0.07$ & --- & --- \\
         &  $6$ & $0.73$ &    $28.0$ & $42.3$ & $64.1$ & $\phm2.41$ &    $-0.55$ & $\phm1.59$ & $\phm0.33$ & --- & --- \\
         &  $7$ & $0.38$ &    $45.2$ & $57.5$ & $65.9$ & $\phm3.36$ &    $-0.29$ & $\phm2.41$ & $\phm1.05$ & --- & --- \\
\hline
         &  $1$ & $1.60$ & $\phz4.7$ & $\phz9.9$ & $25.4$ &    $-0.70$ &    $-0.56$ &    $-0.20$ &    $-0.90$ &    $-0.26$ &    $-0.93$ \\  
         &  $2$ & $1.55$ & $\phz6.8$ &    $13.7$ & $32.0$ &    $-0.23$ &    $-0.65$ &    $-0.08$ &    $-0.78$ &    $-0.37$ &    $-1.07$ \\
         &  $3$ & $1.38$ & $\phz9.1$ &    $16.8$ & $34.8$ & $\phm0.21$ &    $-0.66$ & $\phm0.18$ &    $-0.56$ &    $-0.29$ &    $-1.01$ \\
$0.5176$ &  $4$ & $1.16$ &    $11.9$ &    $20.1$ & $36.6$ & $\phm0.66$ &    $-0.62$ & $\phm0.50$ &    $-0.31$ &    $-0.15$ &    $-0.90$ \\
         &  $5$ & $0.90$ &    $15.6$ &    $24.1$ & $37.9$ & $\phm1.17$ &    $-0.53$ & $\phm0.89$ & $\phm0.00$ & $\phm0.06$ &    $-0.74$ \\
         &  $6$ & $0.63$ &    $21.5$ &    $29.7$ & $39.2$ & $\phm1.81$ &    $-0.35$ & $\phm1.42$ & $\phm0.44$ & $\phm0.38$ &    $-0.48$ \\
         &  $7$ & $0.33$ &    $34.0$ &    $40.8$ & $40.5$ & $\phm2.77$ & $\phm0.10$ & $\phm2.25$ & $\phm1.24$ & $\phm1.02$ & $\phm0.11$ \\
\hline
         &  $1$ & $1.48$ & $\phz4.2$ & $\phz7.7$ & $19.3$ &    $-0.77$ &    $-0.61$ &    $-0.33$ &    $-0.92$ &    $-0.41$ &    $-0.98$ \\
         &  $2$ & $1.40$ & $\phz6.0$ &    $11.0$ & $24.7$ &    $-0.38$ &    $-0.69$ &    $-0.20$ &    $-0.76$ &    $-0.46$ &    $-1.01$ \\  
         &  $3$ & $1.24$ & $\phz8.0$ &    $13.7$ & $26.9$ &    $-0.00$ &    $-0.65$ & $\phm0.05$ &    $-0.54$ &    $-0.34$ &    $-0.90$ \\
$0.6180$ &  $4$ & $1.04$ &    $10.3$ &    $16.5$ & $28.4$ & $\phm0.39$ &    $-0.57$ & $\phm0.35$ &    $-0.28$ &    $-0.17$ &    $-0.76$ \\
         &  $5$ & $0.81$ &    $13.4$ &    $19.8$ & $29.5$ & $\phm0.85$ &    $-0.43$ & $\phm0.72$ & $\phm0.03$ & $\phm0.07$ &    $-0.56$ \\
         &  $6$ & $0.56$ &    $18.3$ &    $24.6$ & $30.5$ & $\phm1.45$ &    $-0.18$ & $\phm1.23$ & $\phm0.49$ & $\phm0.43$ &    $-0.24$ \\
         &  $7$ & $0.29$ &    $28.7$ &    $33.9$ & $31.6$ & $\phm2.40$ & $\phm0.38$ & $\phm2.07$ & $\phm1.31$ & $\phm1.14$ & $\phm0.44$ \\
\hline
\end{tabular*}
\end{small}
\caption{Sample values of the diffusion coefficients in~(\ref{eq:Brusselator2}) 
and the resulting quadratic coefficients in~(\ref{eq:quadraticODEs}),
with $A=3$, $B=9$, $\beta=1$, $r_1=0$ and $r_q=0$, for different choices of
$q$ and~$\alpha$. 
The data are illustrated in Figures~\ref{fig:linear} and~\ref{fig:wnlt}.
A fuller version of this table (for $0.25\leq q\leq0.66$) is available
in full in~\cite{DataRepository}.}
 \label{table:wnlt}
\endgroup\end{table}

Examples of the six quadratic coefficients as functions of~$\alpha$ are shown
in Figure~\ref{fig:wnlt}, for $q=\sqrt{2-\sqrt{3}}=0.5176$, and for $\beta=1$,
with numerical values for this and other choices of~$q$
given in Table~\ref{table:wnlt} and in~\cite{DataRepository}.

With this choice of parameters, the coefficients $\Qzw$ and $\Qzz$ have
opposite sign for $2.33<\alpha<5.00$, and $\Qwz$ and $\Qww$ have opposite sign
for $4.76<\alpha<6.87$. The behaviour of the quadratic coefficients for other
values of~$q$ in the range $0.25\leq q\leq0.66$ is similar: there is a range
of~$\alpha$ for which $\Qzw\Qzz<0$, and (provided $q>\frac{1}{2}$) there is a
range of~$\alpha$ for which $\Qwz\Qww<0$, where the ordering is the same
throughout. The two ranges overlap over a limited range of~$\alpha$, centred on
$\alpha\approx4.8$ for all~$q$. 

\begin{table}
\centering
\begin{tabular}{c|c|c|l}
\hline        
$q$ & $\theta_z$ & $\theta_w$ & Comment\\
\hline
$0.2500$ &  $165.6^\circ$ &  ---         &  \\
$0.3300$ &  $161.0^\circ$ &  ---         &  \\
$0.3780$ &  $158.2^\circ$ &  ---         & $q=1/\sqrt{7}$: superlattice patterns \\
$0.4400$ &  $154.6^\circ$ &  ---         &  \\
$0.5176$ &  $150.0^\circ$ & $30.0^\circ$ & $q=\sqrt{2-\sqrt{3}}$: twelve-fold quasipatterns \\
$0.5500$ &  $148.1^\circ$ & $49.2^\circ$ &  \\
$0.5774$ &  $146.4^\circ$ & $60.0^\circ$ & $q=1/\sqrt{3}$: hexagons \\
$0.6180$ &  $144.0^\circ$ & $72.0^\circ$ & $q=\frac{1}{2}(-1+\sqrt{5})$: ten-fold quasipatterns \\
$0.6600$ &  $141.5^\circ$ & $81.5^\circ$ &  \\
\hline
\end{tabular}
\caption{Values of the length scale ratio~$q$ used in our survey. The angles~$\theta_z$ and $\theta_w$
are defined in Figure~\ref{fig:TWI} and Eq.~(\ref{eq:thetazw}).}
 \label{table:thetas}
\end{table}

\section{Numerical results}

\label{sec:numerical}

Based on the linear and weakly nonlinear calculations in the previous sections, we have
carried out a series of numerical simulations of the PDEs in (\ref{eq:Brusselator2}). Our
main goal is to explore the effect of varying the ratio of length scales,~$q$,
in regimes where we can control the signs of the quadratic coefficients. Our choice
is to fix the diffusive coupling 
coefficient~$\beta=1$ and vary~$\alpha$ with $1\leq\alpha\leq7$ (in steps of 1).
With different choices of~$\alpha$, the two pairs of quadratic coefficients can have the 
same or opposite signs (see Figure~\ref{fig:wnlt}), though the range where both pairs had 
opposite sign was very limited. We chose some special values of~$q$, some less than 
and some greater than~$\frac{1}{2}$:
 $q=1/\sqrt{7}=0.3780$, to encourage superlattice patterns~\cite{DioSilSke1997};
 $q=\sqrt{2-\sqrt{3}}=0.5176$, to encourage twelve-fold quasipatterns~\cite{Mul1994,LifPet1997};
 $q=1/\sqrt{3}=0.5774$, to allow quadratic interactions between six modes on each circle;
 and
 $q=\frac{1}{2}(-1+\sqrt{5})=0.6180$, to encourage ten-fold 
quasipatterns~\cite{FriSon1995}.
We also chose more ``generic'' values of~$q$: $0.25$, $0.33$, $0.44$, $0.55$ 
and~$0.66$. The values of~$q$ and the corresponding angles~$\theta_z$ and $\theta_w$ are 
listed in Table~\ref{table:thetas}.
All chemical properties are frozen with the
choice of $A=3$ and $B=9$ as in~\cite{YanDolZha2002}.

The values of the diffusion coefficients at the codimension-two point 
$r_1=r_q=0$ are given
in Figure~\ref{fig:linear} and in~\cite{DataRepository}. For each selected case of $q$
and~$\alpha$, we vary the diffusion coefficients to explore small positive and
negative values of the two growth rates~$r_1$ and~$r_q$. Specifically, setting
$(r_1,r_q)=(r\cos\theta,r\sin\theta)$, we choose $r=0.01$ (apart from data in
Figure~\ref{fig:stc003}), with $\theta$~varying from $5^{\circ}$ to $355^{\circ}$ in
steps of~$10^{\circ}$. For smaller~$q$ and~$\alpha$, these choices lead to growth
rates~$\sigma(k)$ that are sharply peaked at $k=q$ and $k=1$, with a relatively 
deep negative
minimum in between (see Figure~\ref{fig:dispersion}). However, for larger $q$
and~$\alpha$, the minimum between the two maxima is quite shallow, which means that, even
with a small value of $r=0.01$, there can be wide bands of unstable wavenumbers.

We start all simulations from small-amplitude random initial conditions in
$16\pi\times16\pi$ ($8\times8$ of the shorter wavelengths) domains, except in the case when
$q=\sqrt{2-\sqrt{3}}$ where we also start simulations from a small-amplitude quasipattern
initial condition. For
parameter choices that do not result in a simple pattern, 
we explore the effect of a larger domain by re-running
calculations in $60\pi\times60\pi$ ($30\times30$~wavelengths) domains.
Both $8\times8$ and $30\times30$ domains are
appropriate for twelve-fold quasipatterns~\cite{RucSil2009}. Time simulations
are for at least 10000 time units: this is 100~growth times (for
$r=0.01$) and approximately 
three diffusion times for the larger domain when considering the smallest
values of the diffusion coefficients. 

We use $128\times128$ Fourier modes (using FFTW~\citep{FriJoh2005}, the fastest Fourier 
transform in the West) in each direction
for the $8\times8$ domains, and $512\times512$ Fourier modes for the larger $30\times30$
domains. We use the second-order exponential time differencing
(ETD2)~\citep{CoxMat2002} scheme for timestepping, 
with a fixed timestep of~0.01. For this
matrix exponential method, 
we split the 
linear part of the PDE~(\ref{eq:Brusselator2}) 
into diagonal and off-diagonal parts, and we treat the off-diagonal parts as
nonlinear terms.

In all, we carried out over 4000 simulations, and the results we present below are an
overview of the range of patterns we find. 
For $\alpha\leq3$, we find a wide range of different patterns, but for
$\alpha\geq4$ we find simple patterns (hexagons) almost exclusively.
Therefore, we focus on the cases with $\alpha=1$, 2 and~3. When $\alpha=1$,
all quadratic coefficients are negative for all~$q$
(see Figure~\ref{fig:wnlt}, Table~\ref{table:wnlt}
and~\cite{DataRepository}). Therefore, from Table~\ref{table:predictions}, 
we expect to find only
steady patterns. For $\alpha=2$, $\Qzw$ and $\Qzz$ are of opposite sign
for $q\in\{0.2500,0.3300,0.3780\}$ and are of the same sign for
$q\in\{0.4400,0.5176,0.5774,0.6180,0.6600\}$, although $\Qzz$ is very close to zero for
$q=0.4400$. For $\alpha=3$, $\Qzw$ and $\Qzz$ are of opposite sign for all~$q$ apart
from $q=0.6600$. 
For $1\leq\alpha\leq3$, $\Qwz$ and~$\Qww$ are both negative.
We connect some of the observed steady patterns in this section to the
three cases of nonlinear wave-vector interactions described in 
Figure~\ref{fig:threeexamples} and relate these to our expectations in 
Table~\ref{table:predictions}.

\begin{figure}
\centering
\hbox to \textwidth{\hfil%
  \hbox to 0.35\textwidth{\hfil\textit{(a) $q=0.2500$}\hfil}\hfil
  \hbox to 0.35\textwidth{\hfil\textit{(b) $q=0.3300$}\hfil}\hfil}
\vspace{-3.0ex}
\hbox to \textwidth{\hfil%
\includegraphics[angle=270,width=0.35\textwidth]{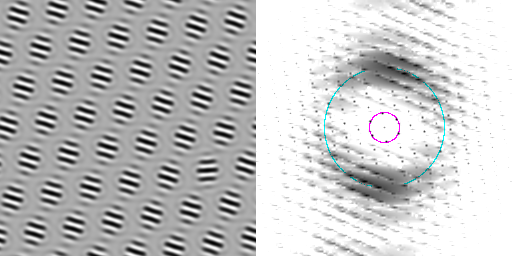}\hfil   
\includegraphics[angle=270,width=0.35\textwidth]{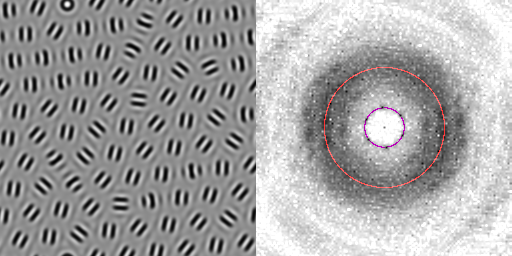}\hfil}
\vspace{1.5ex}
\hbox to \textwidth{\hfil%
  \hbox to 0.35\textwidth{\hfil\textit{(c) $q=0.3780$}\hfil}\hfil
  \hbox to 0.35\textwidth{\hfil\textit{(d) $q=0.4400$}\hfil}\hfil}
\vspace{-3.0ex}
\hbox to \textwidth{\hfil%
\includegraphics[angle=270,width=0.35\textwidth]{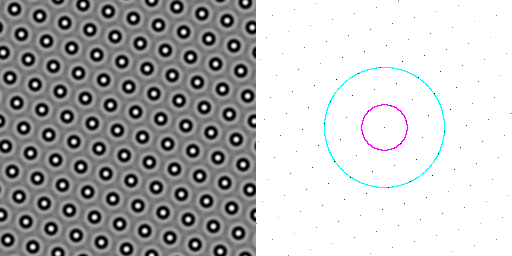}\hfil
\includegraphics[angle=270,width=0.35\textwidth]{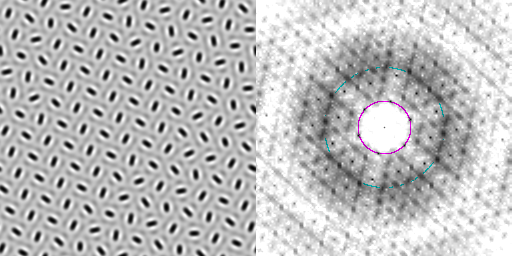}\hfil}
\caption{Examples of patterns, all in $30\times30$ domains for $\alpha=1$ and $r=0.01$, $\theta=45^\circ$,
 with $q$ running from $0.25$ to $0.66$. Each image has a grey scale representing~$U_1(\bx)$
 (the scaling is different in each case)
 and a power spectrum with circles
 $k=1$ and $k=q$ indicated.}
\label{fig:alpha1egs}
\end{figure}

\addtocounter{figure}{-1} 

\begin{figure}
\centering
\hbox to \textwidth{\hfil%
  \hbox to 0.35\textwidth{\hfil\textit{(e) $q=0.5176$}\hfil}\hfil
  \hbox to 0.35\textwidth{\hfil\textit{(f) $q=0.5774$}\hfil}\hfil}
\vspace{-3.0ex}
\hbox to \textwidth{\hfil%
\includegraphics[angle=270,width=0.35\textwidth]{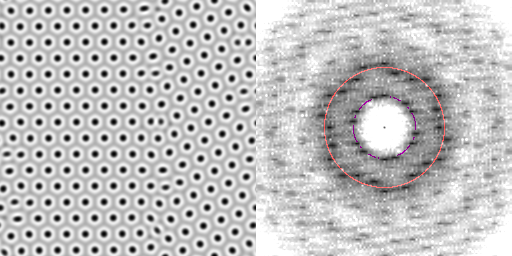}\hfil   
\includegraphics[angle=270,width=0.35\textwidth]{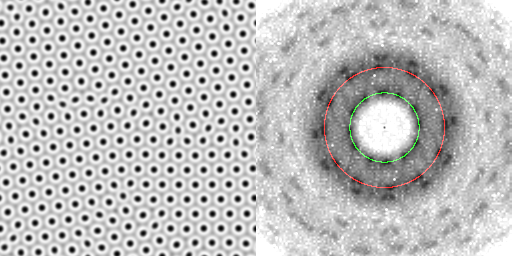}\hfil}
\hbox to \textwidth{\hfil%
  \hbox to 0.35\textwidth{\hfil\textit{(g) $q=0.6180$}\hfil}\hfil
  \hbox to 0.35\textwidth{\hfil\textit{(h) $q=0.6600$}\hfil}\hfil}
\vspace{-3.0ex}
\hbox to \textwidth{\hfil%
\includegraphics[angle=270,width=0.35\textwidth]{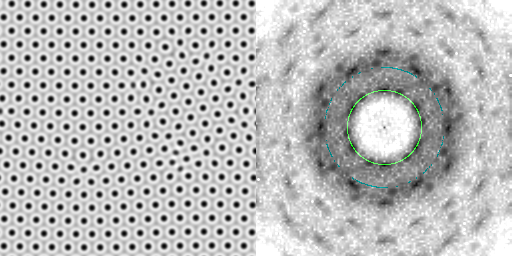}\hfil
\includegraphics[angle=270,width=0.35\textwidth]{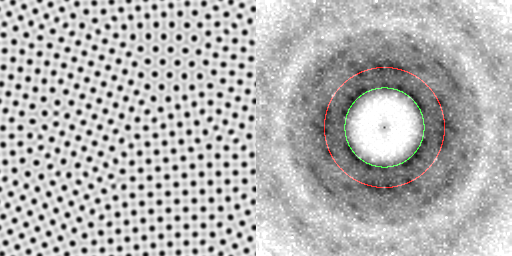}\hfil}
\caption{Continued from previous page.}
\label{fig:alpha1egs}
\end{figure}

\subsection{Steady patterns with varying~$q$: $\alpha=1$}

First, we explore steady patterns with $\alpha=1$ at fixed $r=0.01$ and 
$\theta=45^\circ$,
so $r_1=r_q=0.00707$, but for varying~$q$ (see Figure~\ref{fig:alpha1egs}). 
For $q<\frac{1}{2}$, 
we see strong hexagonal motifs on a scale set by the smaller wavenumber~$q$, inset
with stripes on a scale of wavenumber~1, resembling patterns found
by~\cite{YanDolZha2002}. For $q=0.3780$ the pattern is exactly hexagonal, with six
equally spaced modes on the inner circle and twelve unequally spaced on the outer, as in
Figure~\ref{fig:threeexamples}(a) -- this is the simplest example of a superlattice
pattern. As $q$ increases beyond 0.5176, the patterns continue as essentially
hexagonal on the scale of the smaller wavenumber, but defects and grain boundaries become
more common for larger~$q$.

\begin{figure}
\centering
\hbox to \textwidth{\hfil%
  \hbox to 0.45\textwidth{\hfil\textit{(a) $\alpha=1$, $q=0.5176$}\hfil}\hfil
  \hbox to 0.45\textwidth{\hfil\textit{(b) $\alpha=2$, $q=0.2500$}\hfil}\hfil}
\vspace{0.5ex}
\hbox to \textwidth{\hfil%
\includegraphics[angle=270,width=0.45\textwidth]{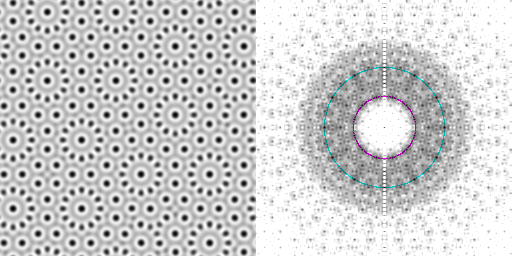}\hfil
\includegraphics[angle=270,width=0.45\textwidth]{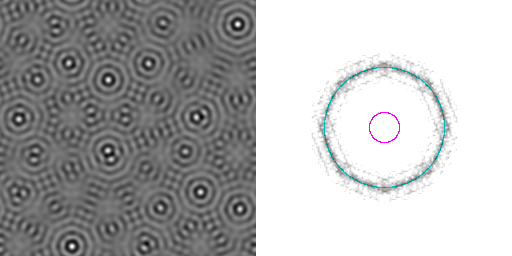}\hfil}
\caption{Examples of quasipatterns. 
 (a)~A twelve-fold             quasipattern with $\alpha=1$, $q=0.5176$, $\theta=315^\circ$.
 (b)~An eight-fold approximate quasipattern with $\alpha=2$, $q=0.2500$, $\theta=165^\circ$.}
\label{fig:qpegs}
\end{figure}

\subsection{Quasipatterns}

We take
$q=\sqrt{2-\sqrt{3}}=0.5176$ and start with small amplitudes for
twelve Fourier modes on the circle $k=1$ as initial condition
to encourage
twelve-fold quasipatterns, finding stable examples as in Figure~\ref{fig:qpegs}(a). This is
a periodic approximant to a true quasipattern, but 
the approximation is particularly accurate in the 
$30\times30$ domain~\cite{RucSil2009}. There are
twelve peaks on the inner and outer circles, interleaved as in
Figure~\ref{fig:threeexamples}(b). This kind of quasipattern has been seen in many
similar kinds of calculations going back to~\cite{Mul1994,LifPet1997}.

We also obtain an eight-fold quasipattern, in Figure~\ref{fig:qpegs}(b). This is
surprising since neither $\theta_z$ nor $\theta_w$ is a multiple of~$45^\circ$ 
(Table~\ref{table:thetas}). In
addition, in our $30\times30$ domain, the approximation to a true eight-fold 
quasipattern is not
particularly accurate.
Nonetheless, there are eight reasonably clear peaks on the inner
circle, with sixteen diffuse peaks on the outer and an additional eight peaks just outside the
outer circle, giving the impression of a regular octagon. It may be 
significant that
$\theta_z=165.6^\circ$ (see Table~\ref{table:thetas}), which is close to
$15^\circ$ less than~$180^\circ$, as the twenty-four peaks on and just off the outer circle are
spaced roughly $15^\circ$ apart.

\begin{figure}
\centering
\hbox to \textwidth{%
  \hbox to 0.45\textwidth{\hfil\textit{(a) $\alpha=2$, $q=0.5176$}\hfil}\hfil
  \hbox to 0.45\textwidth{\hfil\textit{(b) $\alpha=2$, $q=0.6180$}\hfil}}
\vspace{0.5ex}
\hbox to \textwidth{%
\includegraphics[angle=270,width=0.45\textwidth]{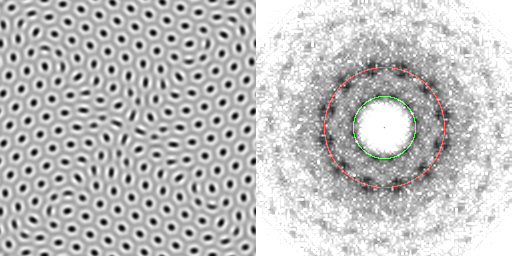}\hfil   
\includegraphics[angle=270,width=0.45\textwidth]{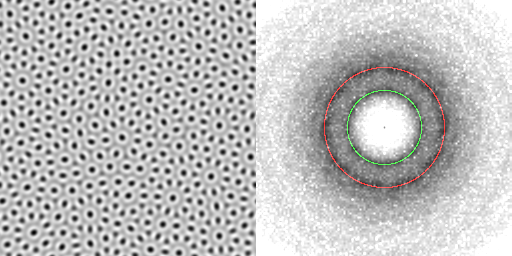}}
\caption{Examples of steady (or persistent) complex patterns. 
 (a)~Swirly distorted hexagonss with $\alpha=2$, $q=0.5176$, $\theta=45^\circ$.
 After a transient of about 20000 time units, these are replaced by hexagons. 
 (b)~Hints of ten-fold quasipattern motifs with $\alpha=2$, $q=0.6180$, $\theta=275^\circ$.
 This complex pattern persists for at least 50000 time units.}
\label{fig:complextegs}
\end{figure}

The third common two-dimensional quasipattern has ten-fold symmetry. We have not
found examples of such a quasipattern, but there are hints of a ten-fold motif in
calculations with $q=0.6180$ (see Figure~\ref{fig:complextegs}b).

\subsection{Steady complex patterns}

We find many examples of hexagonal patterns with defects as in
Figure~\ref{fig:alpha1egs}\hbox{(e--h)}. 
In Figure~\ref{fig:complextegs}, we show two examples
of steady complex patterns that are not just straightforward 
patches of hexagons (as in Figure~\ref{fig:alpha1egs}e--h). In
Figure~\ref{fig:complextegs}(a), with $q=0.5176$, the pattern has a ``swirly'' appearance
with regions of 
distorted hexagons in between patches of more regular hexagons. The patches are
rotated with respect to each other, leading to twelve broad peaks in the outer circle of
the power spectrum. In Figure~\ref{fig:complextegs}(b), with $q=0.6180$, the complex
structure of the pattern is more uniformly distributed, both in space and around the two
circles in the power spectrum. There are several examples of a ten-fold motif, not
surprising given that $q$ is the inverse of the golden ratio.

Both examples are not steady but continue to evolve on timescales longer than 10000 time 
units. The example in Figure~\ref{fig:complextegs}(a) eventually anneals to hexagons. 
The example in Figure~\ref{fig:complextegs}(b) persists for at least 
50000 time units, and
is the closest we have
found to an example of a steady complex pattern with 
the infinite set of wavevectors implied by
Figure~\ref{fig:threeexamples}(c).

\begin{figure}
\centering
\hbox to \textwidth{\hfil \textit{(a) $\alpha=2$, $q=0.4400$}\hfil}
\vspace{-3.0ex}
\hbox to \textwidth{\hfil%
\includegraphics[angle=270,width=0.320\textwidth]{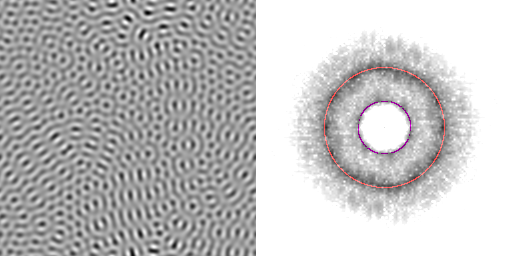}\hfil   
\includegraphics[angle=270,width=0.320\textwidth]{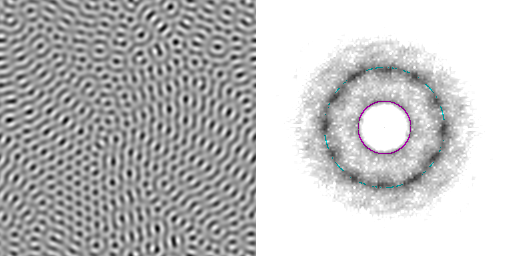}\hfil   
\includegraphics[angle=270,width=0.320\textwidth]{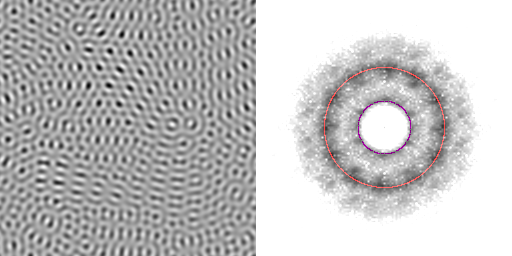}\hfil}   
\hbox to \textwidth{\hfil%
\includegraphics[angle=270,width=0.320\textwidth]{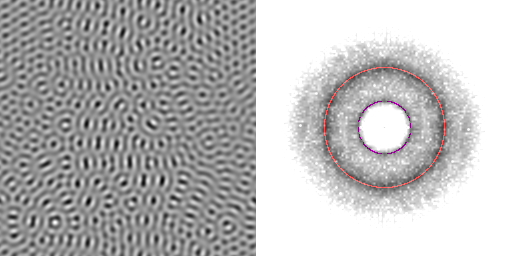}\hfil   
\includegraphics[angle=270,width=0.320\textwidth]{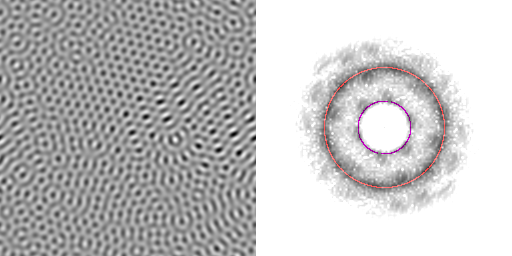}\hfil   
\includegraphics[angle=270,width=0.320\textwidth]{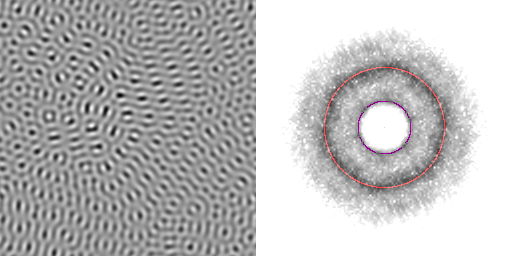}\hfil}
\caption{Frames from two examples of spatiotemporal chaos. (a)~$\alpha=2$, $q=0.4400$, $\theta=5^\circ$.
(b)~$\alpha=3$, $q=0.6180$, $\theta=45^\circ$. The time interval between
frames is 2000~time units.
Videos are available in~\cite{DataRepository}.}
\label{fig:stc}
\end{figure}

\addtocounter{figure}{-1} 

\begin{figure}
\centering
\hbox to \textwidth{\hfil \textit{(b) $\alpha=3$, $q=0.6180$}\hfil}
\vspace{-3.0ex}
\hbox to \textwidth{\hfil%
\includegraphics[angle=270,width=0.320\textwidth]{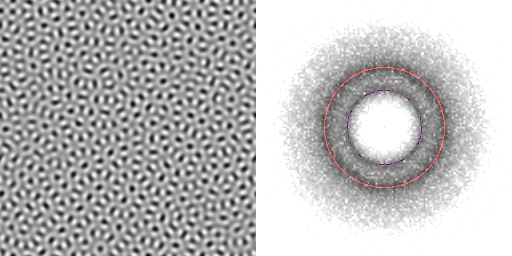}\hfil   
\includegraphics[angle=270,width=0.320\textwidth]{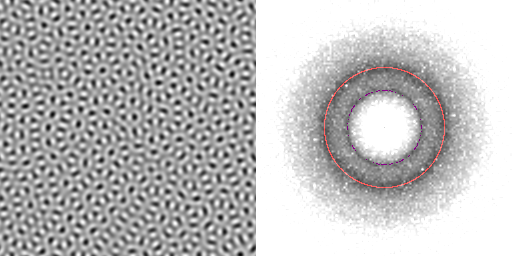}\hfil   
\includegraphics[angle=270,width=0.320\textwidth]{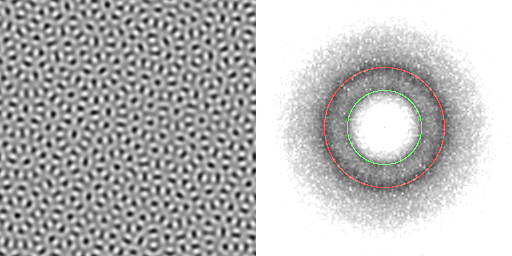}\hfil}
\hbox to \textwidth{\hfil%
\includegraphics[angle=270,width=0.320\textwidth]{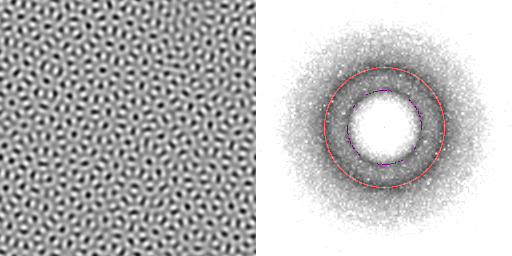}\hfil   
\includegraphics[angle=270,width=0.320\textwidth]{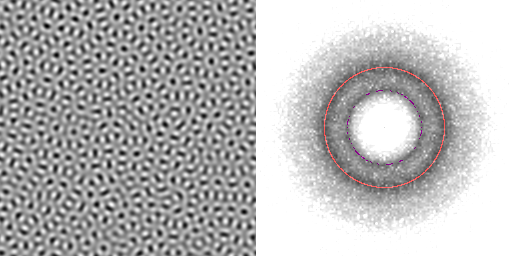}\hfil   
\includegraphics[angle=270,width=0.320\textwidth]{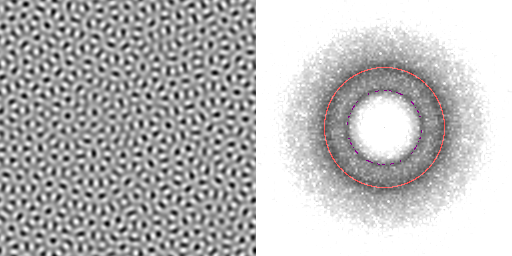}\hfil}
\caption{Continued from previous page.}
\end{figure}


\begin{figure}
\centering
\hbox to \textwidth{\hfil $\alpha=3$, $q=0.3780$\hfil}
\vspace{-3.0ex}
\hbox to \textwidth{\hfil%
\includegraphics[angle=270,width=0.320\textwidth]{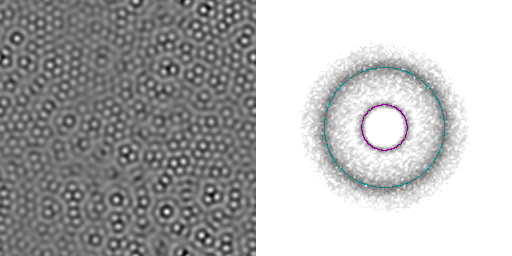}\hfil   
\includegraphics[angle=270,width=0.320\textwidth]{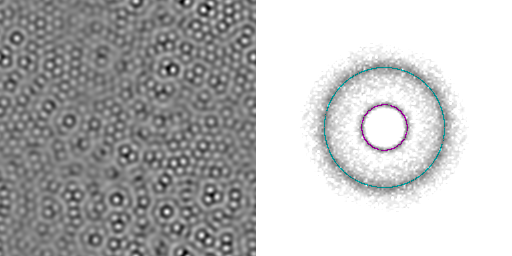}\hfil   
\includegraphics[angle=270,width=0.320\textwidth]{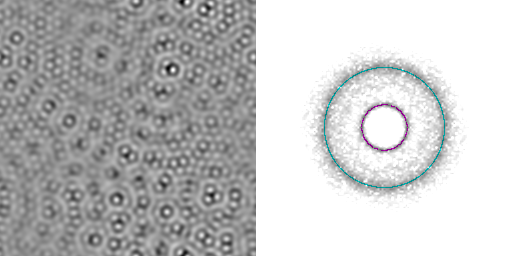}\hfil}   
\hbox to \textwidth{\hfil%
\includegraphics[angle=270,width=0.320\textwidth]{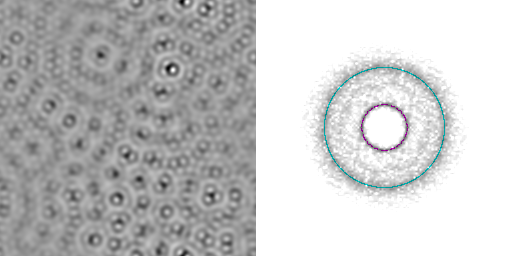}\hfil   
\includegraphics[angle=270,width=0.320\textwidth]{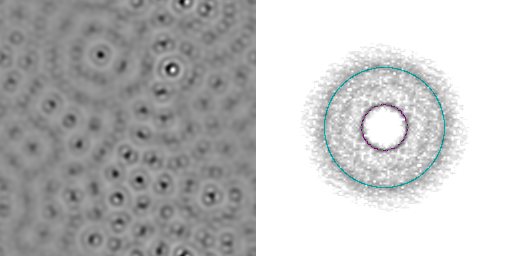}\hfil   
\includegraphics[angle=270,width=0.320\textwidth]{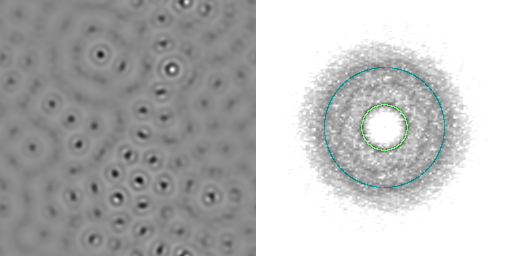}\hfil}
\caption{Frames from an examples of spatiotemporal chaos:
$\alpha=3$, $q=0.3780$, $r=0.03$, $\theta=145^\circ$. 
This example evolves much more quickly than those of Figure~\ref{fig:stc}: 
the time interval between frames is 20~time units.
A video is available in~\cite{DataRepository}.}
\label{fig:stc003}
\end{figure}

\subsection{Spatiotemporal chaos}

Finally, we show three examples of spatiotemporal chaos in Figure~\ref{fig:stc}(a)
and~(b), with $q=0.4400$ and $q=0.6180$ respectively, and in Figure~\ref{fig:stc003},
with $q=0.3780$ but with larger linear parameters than the others calculations
($r=0.03$). The spatiotemporal chaos examples in Figure~\ref{fig:stc} evolve quite slowly: a frame spacing
of 2000~time units is needed to show appreciable differences between the frames. The
frame spacing in Figure~\ref{fig:stc003} is 100 times less. Videos of all three examples
are available in~\cite{DataRepository}.

In the $q<\frac{1}{2}$ example in Figure~\ref{fig:stc}(a), there are evolving
patches of elongated hexagons, and in some frames, the power spectrum has twelve peaks on the
outer circle. In contrast, the $q>\frac{1}{2}$ example in Figure~\ref{fig:stc}(b) has a
much more axisymmetric power spectrum and the complexity of the pattern is more
uniformly spread across the domain. In the third example in Figure~\ref{fig:stc003}, the
system alternates between episodes dominated by small hexagons and episodes dominated by
larger structures.

\section{Summary and Discussion}

\label{sec:discussion}

%

One main finding is that we only find persistent time dependence (as opposed to slow
healing of defects and coarsening of grain boundaries) when $\Qzw$ and $\Qzz$ had
opposite sign, as in Figures~\ref{fig:stc} and~\ref{fig:stc003}. This is consistent with
our \emph{a priori} expectations outlined in Table~\ref{table:predictions}. With
appropriate initial conditions, we find twelve-fold quasipatterns only in the case with
$q=\sqrt{2-\sqrt{3}}$, although we find eight-fold and hints of ten-fold quasipatterns
in other cases. We did find examples of steady (or persistent) 
complex patterns, as in Figure~\ref{fig:complextegs}. It was noticeable that
having two interacting wavelengths encourages patterns with defects.

With quadratic coefficients $\Qzw$ and $\Qzz$ of opposite sign, the preliminary
$8\times8$ calculations often yield time-dependent patterns, with relatively simple
oscillatory, chaotic or heteroclinic cycle dynamics. When we extend these into $30\times30$ domains, the
simple time dependence is often replaced by spatiotemporal chaos, supporting the
infinite set of wavevectors picture implied by Figure~\ref{fig:threeexamples}(c). We
suspect that the reason for this is that in $8\times8$ domains, there are relatively few
modes available close enough to each circle to participate in the dynamics. In contrast,
with $30\times30$ domains, the density of modes in Fourier space is higher, and so modes
are more likely to be able to participate in multiple three-wave interactions, as in
Figure~\ref{fig:TWI}. Considering a single set of modes coupled by a three-wave interaction, the
modes may be oscillatory. When two (or more) sets of modes, also coupled within themselves by 
three-wave interaction, have modes in common, the common
modes will be torn in different directions by their partners in the different sets, resulting in
spatiotemporal chaos.

All interesting cases of time dependence have $\Qzw$ and $\Qzz$ of opposite
sign, and time dependence can happen 
for all values of~$q$. Having $\Qwz$ and $\Qww$ of opposite sign
(relevant only for $q>\frac{1}{2}$) did not lead to persistent time dependence. Having
$q>\frac{1}{2}$ did appear to help when seeking steady complex patterns
(Figure~\ref{fig:complextegs}b): we find no examples of such patterns with
$q<\frac{1}{2}$. This is in contrast to the hypotheses of~\cite{RucSilSke2012}, who
argued that having $q>\frac{1}{2}$ and having $\Qwz$ and $\Qww$ of opposite sign should
encourage complex patterns. In fact, we find that $q<\frac{1}{2}$ is more interesting
than anticipated from the results of~\cite{RucSilSke2012}, especially when $\Qzw$ and
$\Qzz$ have opposite sign: there are many more states possible, including spatiotemporal
chaos, going well beyond the steady superlattice example associated with
$q=1/\sqrt{7}=0.3780$, in Figure~\ref{fig:alpha1egs}(c). We also had not anticipated 
finding quasipatterns in the case $q<\frac{1}{2}$ (as in Figure~\ref{fig:qpegs}b), but
recent work~\cite{IooRuc2020} suggests this warrants more exploration.

In summary, our main numerical findings described above are broadly in line with the
\emph{a priori} expectations in Table~\ref{table:predictions}. In particular, time
dependence, and with it complex spatial structure, requires $\Qzw$ and $\Qzz$ to have
opposite sign. The other pair of quadratic coefficients ($\Qwz$ and $\Qww$) can also have
opposite sign for $\alpha=5$ and~6, but for these values, we mainly find domains of
hexagons. 
{\red At this point, it is not clear why this happens, nor whether other systems
would behave differently.}
We should emphasise that the time-dependence we have found is not associated
with a primary Hopf bifurcation to spiral waves, common in many Turing systems.

It would be interesting to explore these complex patterns in more detail,
in the context of coupled reaction--diffusion systems,
in the context of amplitude equations, especially in the case
$q<\frac{1}{2}$, as initiated in~\cite{Riy2012}, and in the context of simpler model
PDEs, such as the ones proposed by~\cite{Mul1994,LifPet1997,RucSilSke2012,RucSil2009} as models for Faraday waves. 
A related PDE is known to produce 
three-dimensional icosahedral quasipatterns~\cite{SubArcKno2016} and localised
quasipatterns~\cite{SubArcKno2018}, and such structures may also be possible in Turing 
systems. We plan to undertake further investigations in the
future.

Of course, one has to ask whether these kinds of patterns can be found in experiments,
and indeed if the mechanisms for forming them are as outlined in this paper. As explained
in~\cite{BerDolYan2004}, manipulating the strength of the coupling, and indeed the
diffusion constants, as we have done here, is difficult. Nonetheless, the spatially
complex experimental patterns reported in~\cite{BerDolYan2004} (see 
Figure~\ref{fig:Berenstein}) resemble, at least
qualitatively, the images in~Figures~\ref{fig:stc} and~\ref{fig:stc003}.

\section*{Acknowledgements}

We are grateful for conversations with
Mary Silber, G{\'e}rard Iooss, Andrew Archer and
Tomonari Dotera. 
We thank Irving Epstein for permission to reproduce Figure~\ref{fig:Berenstein} 
from~\cite{BerDolYan2004}.
We are also grateful for financial support from the EPSRC: summer
research bursaries (JKC, DJR) and grants number EP/P015689/1 (DJR) and EP/P015611/1
(AMR). AMR is also grateful for support from the Leverhulme Trust (RF-2018-449/9), and PS
is grateful for a L'Or\'eal UK and Ireland Fellowship for Women in Science. CMT is
supported by National Science Foundation grant DMS-1813752.

\bibliography{turing}

\end{document}